\documentclass[aps,twocolumn,prb,amsmath,amssymb,superscriptaddress]{revtex4-2}

\usepackage{url}
\usepackage{graphicx}
\usepackage{color}
\usepackage{natbib}
\usepackage{hyperref}
\usepackage{notes2bib}
\usepackage{mathbbol}
\usepackage{float}
\usepackage{ulem}
\usepackage{tikz}
\usepackage{xcolor}
\usepackage{svg}
\usepackage{ulem}

\def\dg{{\;\dagger}}
\def\doubleunderline#1{\underline{\underline{#1}}}

\begin{document}

\parindent=0pt

\title{Finite-frequency noise, Fano factor, $\Delta T$-noise\\ and cross-correlations in double quantum dots}
\author{A. Cr\'epieux\footnote{Corresponding author}}
\address{Aix Marseille Univ, Universit\'e de Toulon, CNRS, CPT, Marseille, France}
\author{T.Q. Duong}
\address{Aix Marseille Univ, Universit\'e de Toulon, CNRS, CPT, Marseille, France}
\author{M. Lavagna}
\address{Univ. Grenoble Alpes, CEA, IRIG, PHELIQS, F-38000 Grenoble, France}

\begin{abstract}
A theoretical study on electrical current fluctuations in a double quantum dot connected to electronic reservoirs is presented, with the aim of deriving the finite-frequency noise, the Fano factor and the $\Delta T$-noise. We establish a general expression for the noise in terms of Green functions in the double quantum dot and self-energies in the reservoirs. This result is then applied to model double quantum dots in various situations. For a non-interacting double quantum dot, we highlight several interesting features in the physical properties of this system. In particular, we demonstrate the possibility of obtaining a significant reduction in zero-frequency noise and Fano factor either when the system is placed in a given operating regime, or when a temperature gradient is applied between the two reservoirs, resulting in a negative $\Delta T$-noise being generated. In addition, in the vicinity of honeycomb vertices, a sign change is observed in the finite-frequency cross-correlator between the two reservoirs, in contrast to what is obtained for the zero-frequency cross-correlator, which remains negative throughout the $(\varepsilon_1,\varepsilon_2)$-plane, $\varepsilon_{1,2}$ being the level energies in each of the two dots. By using an approximate first-level numerical approach, we finally study how the finite-frequency noise in a double quantum dot evolves under the influence of Coulomb interactions.
\end{abstract}

\maketitle

%%%%%%%%%%%%%%%%%%%%%%%%%%%%%%%%%%%%%%%%%%%%%%%%%%%%%%%%%%%%%%%%%%
%																 %
%																 %
%		INTRODUCTION											     %
%																 %
%																 %
%%%%%%%%%%%%%%%%%%%%%%%%%%%%%%%%%%%%%%%%%%%%%%%%%%%%%%%%%%%%%%%%%%

\section{Introduction}

Reducing electrical noise in double quantum dots is a major challenge if we wish to finely control electrical charge transfer and thus improve the performance and quality factor of these systems, particularly in spin-qubits\cite{Martins2016,Connors2022,PaqueletWuetz2023}. Theoretical studies have been carried out to characterize current fluctuations and electrical noise in double quantum dots\cite{Zenelaj2022,Zou2024}. However, they are mostly limited to the calculation of noise and cross-correlations at zero-frequency \cite{Kiesslich2003,Lopez2004,Wu2005,Aghassi2006,Bodoky2008,Weymann2008,Dong2009,Lu2010,Zhao2012,Luo2013,Lu2012,Lu2014,Lu2016,Zocher2013,Horovitz2019,Maslova2021,Wrzesniewski2024} or finite-frequency noise based on perturbative approaches\cite{Sun2000,Sukhorukov2001,Wu2006,Lambert2007,Marcos2011,Jin2015,Shi2016,Xu2022}, using principally the master equation technique. In addition, a number of experimental studies have investigated current fluctuations and electrical noise in double quantum dots constructed from GaAs/AlGaAs or Si/SiGe heterostructures. This enables to explore not only the zero-frequency case but also the finite-frequency case. This type of experimental studies is set to expand considerably in the years ahead. Some of these works aim to establish the optimum conditions for reducing the device sensitivity to electrical noise, with a view to obtaining long-lived, high-fidelity spin qubits\cite{Martins2016,Connors2022,PaqueletWuetz2023}, while other experimental studies focus on measuring cross-correlations or spatial correlations of electrical currents\cite{McClure2007,Yoneda2023,Cywinski2023}, fluctuation distributions of current\cite{Singh2016,Singh2019a} or entropy\cite{Singh2019b}. Taken together, these experimental works provide the motivation for developing further theoretical studies on electrical noise and current fluctuations in double quantum dots.

In this work,  using the non-equilibrium Green function technique, we perform non-perturbative calculations to establish the general expression for the finite-frequency noise in a double quantum dot system connected to left and right electron reservoirs. We numerically calculate the noise and the Fano factor in different geometries of the double quantum dot, either connected in series or in parallel, and discuss the results. We then turn our attention to the $\Delta T$-noise generated when the two reservoirs are heated to different temperatures\cite{Rech2020,Zhitlukhina2020,Eriksson2021,Hasegawa2021,Rebora2022,Popoff2022,Zhang2022,Zhitlukhina2023,Hubler2023}. We show for the first time that the noise follows a very peculiar characteristic evolution as a function of $\varepsilon_1$ and $\varepsilon_2$, a result that has never been demonstrated before in double quantum dots. We also determine the cross-correlator between the left and right reservoirs, compare the results obtained with experimental results\cite{McClure2007}, and discuss the possible origins of the observed sign change. Whenever possible, we will consider our findings in the light of existing results in the literature.

The paper is organized as follows: the calculation of finite-frequency noise in a double quantum dot is presented in Section~\ref{section2}, with details given in Appendix. The behavior of Fano factor, $\Delta T$-noise and finite-frequency cross-correlator in a non-interacting double quantum dot arranged either in a serial or in a parallel geometry are discussed in Section~\ref{noninteracting}, in particular their evolution in the $(\varepsilon_1,\varepsilon_2)$-plane, where $\varepsilon_{1}$ and $\varepsilon_{2}$ are the energy levels of dots $1$ and $2$. In Section~\ref{interacting}, we study the finite-frequency noise in a Coulomb interacting double quantum dot. We conclude in Section~\ref{conclusion}.

%%%%%%%%%%%%%%%%%%%%%%%%%%%%%%%%%%%%%%%%%%%%%%%%%%%%%%%%%%%%%%%%%%
%																 %
%																 %
%		FINITE-FREQUENCY NOISE											     %
%																 %
%																 %
%%%%%%%%%%%%%%%%%%%%%%%%%%%%%%%%%%%%%%%%%%%%%%%%%%%%%%%%%%%%%%%%%%

\begin{figure}
\begin{center}
\includegraphics[width=8cm]{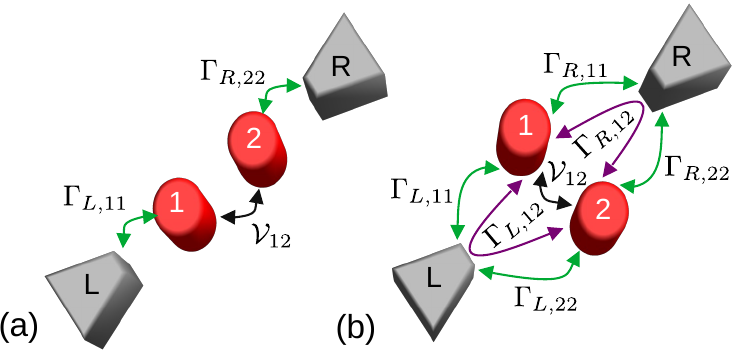}
\vspace*{-0.2cm}
\caption{Schematic picture of a double quantum dot connected to left~(L) and right~(R) reservoirs of electrons, arranged either (a)~in series, or (b)~in parallel. The coupling between the dots 1 and 2 is denoted as $\mathcal{V}_{12}$, while the coupling between reservoir $\alpha=L,R$ and dot $i,j=1,2$ is denoted as $\Gamma_{\alpha,ij}=2\pi\rho_\alpha V_{i\alpha}^*V_{j\alpha}$, where $V_{i\alpha}$ is the hopping integral between dot~$i$ and reservoir~$\alpha$. The density of states, temperature and chemical potential of reservoir~$\alpha$ are denoted $\rho_\alpha$, $T_\alpha$ and $\mu_\alpha$, respectively. The voltage bias is $eV=\mu_L-\mu_R$.}\label{figure0}
\end{center}
\end{figure}

\section{Finite-frequency noise}\label{section2}

\subsection{Definition}

The finite-frequency non-symmetrized noise in a double quantum dot connected to two reservoirs of electrons, as depicted in Fig.~\ref{figure0}, is defined as the Fourier transform of electrical current fluctuations
\begin{eqnarray}\label{def_noise}
\mathcal{S}_{\alpha\beta}(\omega)=\int_{-\infty}^{\infty} \langle \Delta \widehat{I}_\alpha(t) \Delta \widehat{I}_\beta(0) \rangle e^{-i\omega t}dt~,
\end{eqnarray}
where the indices $\alpha$ and $\beta$ refer to left and right reservoirs of electrons, labeled as $L$ and $R$. With this definition, $ \mathcal{S}_{\alpha\alpha}(\omega)$ for negative frequency corresponds to absorption noise, whereas $ \mathcal{S}_{\alpha\alpha}(\omega)$ for positive frequency corresponds to emission noise\cite{Deblock2003,Zamoum2016,Crepieux2018}. The quantities $\mathcal{S}_{\alpha\alpha}(\omega)$ and $\mathcal{S}_{\alpha\overline{\alpha}}(\omega)$ are respectively the auto-correlator and the cross-correlator of the electrical current (with $\overline{\alpha}=R$ when $\alpha=L$, and $\overline{\alpha}=L$ when $\alpha=R$). In Eq.~(\ref{def_noise}), $\Delta \widehat{I}_\alpha(t)$ is the deviation of $\widehat{I}_\alpha(t)$ from its average value~$\langle \widehat I_\alpha\rangle$, where $\widehat{I}_\alpha(t)$ is the current operator in the reservoir $\alpha$, defined as
\begin{eqnarray}\label{definitionI}
 \widehat{I}_\alpha(t)=\frac{ei}{\hbar}\sum_{k\in\alpha}\sum_{i=1,2}\sum_{n\in i}\left(V_{i\alpha}\widehat c_{\alpha k}^\dg\widehat d_{in}-V_{i\alpha}^*\widehat d_{in}^\dg\widehat c_{\alpha k}\right)~,
\end{eqnarray}
where $\widehat c_{\alpha k}^\dg$ ($\widehat c_{\alpha k}$) is the creation (annihilation) operator of one electron in the reservoir $\alpha$, with momentum~$k$, and $\widehat d_{in}^\dg$ ($\widehat d_{in}$) is the creation (annihilation) operator of one electron in the dot~$i$, with $i=1,2$. The index~$n$ corresponds to the orbital degree of freedom in the dot $i$. One assumes that the hopping integral~$V_{i\alpha}$ between the states $|in\rangle$ in the dot~$i$ and $|\alpha k\rangle$ in the reservoir~$\alpha$ depend neither on the orbital $n$, nor on the momentum~$k$ (flat wide-band limit).

By writing Eq.~(\ref{def_noise}) as $ \mathcal{S}_{\alpha\beta}(\omega)=\int_{-\infty}^\infty dt e^{-i\omega t}  \mathcal{S}_{\alpha\beta}(t,0)$, where $  \mathcal{S}_{\alpha\beta}(t,t')=\langle \delta  \widehat I_\alpha(t) \delta  \widehat I_\beta(t') \rangle$, and by using the definition given in Eq.~(\ref{definitionI}), one show that $ \mathcal{S}_{\alpha\beta}(t,t')$ can be expressed as a sum of contributions including four distinct Green functions
\begin{eqnarray}
 \mathcal{S}_{\alpha\beta}(t,t')&=&\frac{e^2}{\hbar^2} \sum_{k\in \alpha,k'\in\beta}
\sum_{i,j=1,2} \sum_{n\in i,m\in j}\nonumber\\
&&\times\Big[ V_{i\alpha}V_{j\beta} G_1^{cd,>}(t,t')-V_{i\alpha}V_{j\beta}^{\ast} G_2^{cd,>}(t,t')\nonumber\\
&&-V_{i\alpha}^{\ast}V_{j\beta} G_3^{cd,>}(t,t') +V_{i\alpha}^{\ast}V_{j\beta}^{\ast} G_4^{cd,>}(t,t')  \Big]\nonumber\\
&&-\langle \widehat I_\alpha \rangle\langle \widehat I_\beta \rangle~,
\end{eqnarray}
with
\begin{eqnarray}
G_1^{cd,>}(t,t')&=&-\langle T\, \widehat c_{\alpha k}^{\;\dag}(t)\widehat d_{in}(t)\widehat c_{\beta k'}^{\;\dag}(t') \widehat d_{jm}(t') \rangle~, \\
G_2^{cd,>}(t,t')&=&-\langle T\,  \widehat c_{\alpha k}^{\;\dag}(t)\widehat d_{in}(t)\widehat d_{jm}^{\;\dag}(t') \widehat c_{\beta k'}(t')  \rangle~, \\
G_3^{cd,>}(t,t')&=&-\langle T\,  \widehat d_{in}^{\;\dag}(t)\widehat c_{\alpha k}(t)\widehat c_{\beta k'}^{\;\dag}(t')\widehat d_{jm}(t')  \rangle~, \\
G_4^{cd,>}(t,t')&=&-\langle T\, \widehat d_{in}^{\;\dag}(t)\widehat c_{\alpha k}(t)\widehat d_{jm}^{\;\dag}(t')\widehat c_{\beta k'}(t')  \rangle~,
\end{eqnarray}
where $T$ is the time ordering operator. We use a S-matrix expansion to express the Green functions $G_\ell^{cd,>}$, with $\ell=1$ to 4, in terms of products of two-particle Green function in the double quantum dot, $G^{dd}$, and Green function in the reservoir $\alpha$, $g_{\alpha k}$ (see \ref{appendixA}), to factorize the two-particle Green functions into a product of two single-particle Green functions (Hartee-Fock level approximation), and next to perform a Fourier transform (see \ref{appendixB}) in order to write the finite-frequency noise in terms of Green function of the double quantum dot and self-energy of the reservoirs (see \ref{appendixC}).

\subsection{Results}

The finite-frequency noise is given by a sum of five contributions, $\mathcal{S}_{\alpha\beta}(\omega)=\sum_{\ell=1}^5\mathcal{P}_{\ell}(\omega)$, with
\begin{eqnarray}\label{noise}
\mathcal{P}_1(\omega)&=&\delta_{\alpha\beta}\frac{e^2}{h}\int_{-\infty}^{\infty} d\varepsilon \mathrm{Tr}\Big\{\Big[
\doubleunderline{\bf G^<}(\varepsilon)\doubleunderline{\mathbb{\Sigma}^>_\alpha}(\varepsilon-\hbar\omega)\nonumber\\
&&+\doubleunderline{\bf G^>}(\varepsilon-\hbar\omega)\doubleunderline{\mathbb{\Sigma}^<_\alpha}(\varepsilon)\Big]\Big\}~,
\end{eqnarray}
\begin{eqnarray}
\mathcal{P}_2(\omega)&=&-\frac{e^2}{h}\int_{-\infty}^{\infty} d\varepsilon \mathrm{Tr}\Big\{\Big[\doubleunderline{{\bf G}^r}(\varepsilon-\hbar\omega)\doubleunderline{\mathbb{\Sigma}^>_\beta}(\varepsilon-\hbar\omega)\nonumber\\
&&+\doubleunderline{\bf G^>}(\varepsilon-\hbar\omega)\doubleunderline{\mathbb{\Sigma}^a_\beta}(\varepsilon-\hbar\omega)\Big]\nonumber\\
&&\times \Big[\doubleunderline{{\bf G}^r}(\varepsilon)\doubleunderline{\mathbb{\Sigma}^<_\alpha}(\varepsilon)+\doubleunderline{\bf G^<}(\varepsilon)\doubleunderline{\mathbb{\Sigma}^a_\alpha}(\varepsilon)\Big]\Big\}~,
\end{eqnarray}
\begin{eqnarray}
\mathcal{P}_3(\omega)&= &\frac{e^2}{h}\int_{-\infty}^{\infty} d\varepsilon \mathrm{Tr}\Big\{\doubleunderline{\bf G^>}(\varepsilon-\hbar\omega)\Big[\doubleunderline{\mathbb{\Sigma}^<_\beta}(\varepsilon)\doubleunderline{{\bf G}^a}(\varepsilon)\doubleunderline{\mathbb{\Sigma}^a_\alpha}(\varepsilon)\nonumber\\
&&+\doubleunderline{\mathbb{\Sigma}^r_\beta}(\varepsilon)\doubleunderline{\bf G^<}(\varepsilon)\doubleunderline{\mathbb{\Sigma}^a_\alpha}(\varepsilon)+\doubleunderline{\mathbb{\Sigma}^r_\beta}(\varepsilon)\doubleunderline{{\bf G}^r}(\varepsilon)\doubleunderline{\mathbb{\Sigma}^<_\alpha}(\varepsilon)\Big]\Big\}~,\nonumber\\
\end{eqnarray}
\begin{eqnarray}
\mathcal{P}_4(\omega)&=&\frac{e^2}{h}\int_{-\infty}^{\infty} d\varepsilon \mathrm{Tr}\Big\{ \doubleunderline{\bf G^<}(\varepsilon)\nonumber\\
&&\times\Big[\doubleunderline{\mathbb{\Sigma}^r_\alpha}(\varepsilon-\hbar\omega)\doubleunderline{{\bf G}^r}(\varepsilon-\hbar\omega)\doubleunderline{\mathbb{\Sigma}^>_\beta}(\varepsilon-\hbar\omega)\nonumber\\
&&+\doubleunderline{\mathbb{\Sigma}^r_\alpha}(\varepsilon-\hbar\omega)\doubleunderline{\bf G^>}(\varepsilon-\hbar\omega)\doubleunderline{\mathbb{\Sigma}^a_\beta}(\varepsilon-\hbar\omega)\nonumber\\
&&+\doubleunderline{\mathbb{\Sigma}^>_\alpha}(\varepsilon-\hbar\omega)\doubleunderline{{\bf G}^a}(\varepsilon-\hbar\omega)\doubleunderline{\mathbb{\Sigma}^a_\beta}(\varepsilon-\hbar\omega)\Big]\Big\}~,
\end{eqnarray}
and,
\begin{eqnarray}
\mathcal{P}_5(\omega)&=&-\frac{e^2}{h}\int_{-\infty}^{\infty} d\varepsilon \mathrm{Tr}\Big\{ \Big[\doubleunderline{\mathbb{\Sigma}^r_\alpha}(\varepsilon-\hbar\omega)\doubleunderline{\bf G^>}(\varepsilon-\hbar\omega)\nonumber\\
&&+\doubleunderline{\mathbb{\Sigma}^>_\alpha}(\varepsilon-\hbar\omega)\doubleunderline{{\bf G}^a}(\varepsilon-\hbar\omega)\Big]\nonumber\\
&&\times \Big[\doubleunderline{\mathbb{\Sigma}^r_\beta}(\varepsilon)\doubleunderline{\bf G^<}(\varepsilon)+\doubleunderline{\mathbb{\Sigma}^<_\beta}(\varepsilon)\doubleunderline{{\bf G}^a}(\varepsilon)\Big]\Big\}~,
\end{eqnarray}
where $\mathrm{Tr}\{\,\}$ corresponds to the trace operator. The matrix elements of the self-energies $\doubleunderline{\mathbb{\Sigma}^{r,a,\lessgtr}_\alpha}(\varepsilon)$ in the reservoir $\alpha$, and the Green functions in the double quantum dot $\doubleunderline{\bf G^{r,a,\lessgtr}}(\varepsilon)$ are respectively defined as
\begin{eqnarray}
 \mathbb{\Sigma}_{\alpha,ij}^{r,a,\lessgtr}(\varepsilon)=\sum_{k\in \alpha} V_{i\alpha}^*V_{j\alpha} g_{k\alpha}^{r,a,\lessgtr}(\varepsilon)~,
\end{eqnarray}
and
\begin{eqnarray}
 {\bf G}_{ij}^{r,a,\lessgtr}(\varepsilon)=\sum_{n\in i,m\in j }G_{in,jm}^{r,a,\lessgtr}(\varepsilon)~.
\end{eqnarray}

The self-energy matrices associated with the reservoir~$\alpha$ are given by $\doubleunderline{\mathbb{\Sigma}_\alpha^{r,a}}(\varepsilon)=\mp (i/2) \doubleunderline{\Gamma_\alpha}$, $\doubleunderline{\mathbb{\Sigma}_\alpha^<}(\varepsilon)=if_\alpha^e(\varepsilon) \doubleunderline{\Gamma_\alpha}$ and $\doubleunderline{\mathbb{\Sigma}_\alpha^>}(\varepsilon)=-if_\alpha^h(\varepsilon) \doubleunderline{\Gamma_\alpha}$, where $f_\alpha^e(\varepsilon)=1/(1+\exp(\varepsilon-\mu_\alpha)/k_BT_\alpha)$ and $f_\alpha^h(\varepsilon) =1-f_\alpha^e(\varepsilon) $ are the Fermi-Dirac distribution functions for electrons and holes respectively, $T_\alpha$ is the temperature and $\mu_\alpha$ the chemical potential of the reservoir $\alpha$ with density of states $\rho_\alpha$, which is energy independent in the flat wide-band limit, and where $\Gamma_{\alpha,ij}=2\pi\rho_\alpha V^*_{i\alpha}V_{j\alpha}$ are the elements of the dot-reservoir coupling matrix $\doubleunderline{\Gamma_\alpha}$. Consequently, we establish in \ref{appendixD} and \ref{appendixE} the following expression for the electrical current auto-correlator $\mathcal{S}_{LL}(\omega)$ in the $L$-reservoir
\begin{eqnarray}\label{SLL}
 \mathcal{S}_{LL}(\omega)&=&\frac{e^2}{h}\int_{-\infty}^{\infty}d\varepsilon\, \mathrm{Tr} \bigg\{f^e_L(\varepsilon)f^h_L(\varepsilon-\hbar\omega)\nonumber\\
&&\times\Big[\doubleunderline{\mathcal{T}_{LL}^\mathrm{eff}}(\varepsilon)\doubleunderline{\mathcal{T}_{LL}^\mathrm{eff}}(\varepsilon-\hbar\omega)\nonumber\\
&& +\left(\doubleunderline{t_{LL}}(\varepsilon)-\doubleunderline{t_{LL}}(\varepsilon-\hbar\omega)\right)\nonumber\\
&&\times\left(\doubleunderline{t^+_{LL}}(\varepsilon)-\doubleunderline{t^+_{LL}}(\varepsilon-\hbar\omega)\right)  \Big]\nonumber\\
 && +f^e_R(\varepsilon)f^h_R(\varepsilon-\hbar\omega)\doubleunderline{\mathcal{T}_{RL}}(\varepsilon)\doubleunderline{\mathcal{T}_{RL}}(\varepsilon-\hbar\omega)\nonumber\\
&&+f^e_L(\varepsilon)f^h_R(\varepsilon-\hbar\omega)\nonumber\\
&&\times\left[1-\doubleunderline{\mathcal{T}_{LL}^\mathrm{eff}}(\varepsilon)\right]\doubleunderline{\mathcal{T}_{RL}}(\varepsilon-\hbar\omega)
\nonumber\\
 && +f^e_R(\varepsilon)f^h_L(\varepsilon-\hbar\omega)\nonumber\\
 &&\times\doubleunderline{\mathcal{T}_{RL}}(\varepsilon)\left[1-\doubleunderline{\mathcal{T}_{LL}^\mathrm{eff}}(\varepsilon-\hbar\omega)\right] \bigg\}~,
\end{eqnarray}
where we have defined the transmission amplitude matrix $\doubleunderline{t_{\alpha\alpha}}(\varepsilon)=i\doubleunderline{{\bf G}^r}(\varepsilon)\doubleunderline{\Gamma_\alpha}$, the transmission coefficient matrix
$ \doubleunderline{\mathcal{T}_{\alpha\beta}}(\varepsilon)=\doubleunderline{{\bf G}^r}(\varepsilon)\,\doubleunderline{\Gamma_\alpha}\;\doubleunderline{{\bf G}^a}(\varepsilon)\,\doubleunderline{\Gamma_\beta}$,
and the effective transmission coefficient matrix $\doubleunderline{\mathcal{T}_{\alpha\alpha}^\mathrm{eff}}(\varepsilon)=\doubleunderline{t_{\alpha\alpha}}(\varepsilon)+\doubleunderline{t_{\alpha\alpha}^+}(\varepsilon)-\doubleunderline{\mathcal{T}_{\alpha\alpha}}(\varepsilon)$,
where $\doubleunderline{t_{\alpha\alpha}^+}(\varepsilon)$ is the conjugate transpose of $\doubleunderline{t_{\alpha\alpha}}(\varepsilon)$.

In a similar way, we find that the expression for the electrical current cross-correlator $\mathcal{S}_{LR}(\omega)$ is given by
 \begin{eqnarray}\label{SLR}
 \mathcal{S}_{LR}(\omega)&=&\frac{e^2}{h}\int_{-\infty}^{\infty}d\varepsilon\, \mathrm{Tr} \bigg\{f^e_L(\varepsilon)f^h_L(\varepsilon-\hbar\omega)\nonumber\\
&&\times\Big[\left[\doubleunderline{\mathcal{T}_{LL}}(\varepsilon)-\doubleunderline{t_{LL}}(\varepsilon)\right]\doubleunderline{\mathcal{T}_{LR}}(\varepsilon-\hbar\omega)\nonumber\\
 && -\doubleunderline{\mathcal{T}_{LL}}(\varepsilon)\doubleunderline{t_{RR}^+}(\varepsilon-\hbar\omega)\Big]\nonumber\\
 && +f^e_R(\varepsilon)f^h_R(\varepsilon-\hbar\omega)\nonumber\\
 &&\times
 \Big[\left[\doubleunderline{\mathcal{T}_{RL}}(\varepsilon)-\doubleunderline{t_{LL}^+}(\varepsilon)\right]\doubleunderline{\mathcal{T}_{RR}}(\varepsilon-\hbar\omega)\nonumber\\
 && -\doubleunderline{\mathcal{T}_{RL}}(\varepsilon)\doubleunderline{t_{RR}}(\varepsilon-\hbar\omega)\Big]
\nonumber\\
 && +f^e_L(\varepsilon)f^h_R(\varepsilon-\hbar\omega)  \left[\doubleunderline{\mathcal{T}_{LL}}(\varepsilon)-\doubleunderline{t_{LL}}(\varepsilon)\right]\nonumber\\
 &&\times
  \left[\doubleunderline{\mathcal{T}_{RR}}(\varepsilon-\hbar\omega)-\doubleunderline{t_{RR}}(\varepsilon-\hbar\omega)\right]
\nonumber\\
 && +f^e_R(\varepsilon)f^h_L(\varepsilon-\hbar\omega)\left[\doubleunderline{\mathcal{T}_{RL}}(\varepsilon)-\doubleunderline{t_{LL}^+}(\varepsilon)\right]\nonumber\\
 &&\times
     \left[\doubleunderline{\mathcal{T}_{LR}}(\varepsilon-\hbar\omega)-\doubleunderline{t_{RR}^+}(\varepsilon-\hbar\omega)\right]
 \bigg\}~.
\end{eqnarray}

The auto-correlator $\mathcal{S}_{RR}(\omega)$ and the cross-correlator $\mathcal{S}_{RL}(\omega)$ are obtained by interchanging the indices~$L$ and~$R$ in Eq.~(\ref{SLL}) and Eq.~(\ref{SLR}), respectively. Both $\mathcal{S}_{LL}(\omega)$  and $\mathcal{S}_{RR}(\omega)$  are real quantities, whereas $\mathcal{S}_{RL}(\omega)$ and $\mathcal{S}_{LR}(\omega)$ take complex values. Eqs.~(\ref{SLL}) and (\ref{SLR}) apply for any geometry of the double quantum dot (serial or parallel) since one can freely play with the values of the dot-reservoir couplings $\Gamma_{\alpha,ij}$ (see Fig.~\ref{figure0}).

The results for the finite-frequency noise given in this Section constitute a generalization to the double quantum dot of the results obtained for a single quantum dot\cite{Zamoum2016,Crepieux2018}. The main differences lie in the expressions themselves of the Green functions and of the transmission amplitudes and coefficients, and, as well, in the presence of matrix products instead of scalar products. The major advance of this result, compared to existing results in the literature on double quantum dots\cite{Sun2000,Lambert2007,Marcos2011,Shi2016,Xu2022}, is that (i)~they are non-perturbative results according to dot-reservoir and dot-dot couplings, (ii)~they are valid at any bias/gate voltage, temperature, and frequency, and (iii)~once the Green function of the double quantum dot is determined, analytically or numerically, it gives direct access to the non-symmetrized finite-frequency noise.

By starting from Eqs.~(\ref{SLL}) and (\ref{SLR}), we present two applications of our results: the noise obtained for a non-interacting double quantum dot in Sec.~\ref{noninteracting}, and for a Coulomb interacting double quantum dot in Sec.~\ref{interacting}, in both serial and parallel geometries.

%%%%%%%%%%%%%%%%%%%%%%%%%%%%%%%%%%%%%%%%%%%%%%%%%%%%%%%%%%%%%%%%%%
%																 %
%																 %
%		NON-INTERACTION DQD											     %
%																 %
%																 %
%%%%%%%%%%%%%%%%%%%%%%%%%%%%%%%%%%%%%%%%%%%%%%%%%%%%%%%%%%%%%%%%%%

\section{Non-interacting double quantum dot}\label{noninteracting}

\subsection{Model}

In the absence of interactions, the Hamiltonian of two coupled quantum dots connected to left~(L) and right~(R) reservoirs, as depicted in Fig.~\ref{figure0}, is given by
\begin{eqnarray}\label{H}
\mathcal{\widehat H}&=&\sum_{\alpha =L,R}\sum_{k \in \alpha} \varepsilon_{\alpha k}^{\,}\widehat c_{\alpha k}^\dg \widehat c_{\alpha k}^{\,} \nonumber\\
&&
+\sum_{i=1,2}\sum_{n \in i} \varepsilon_{in }^{\,}\widehat d_{in}^\dg \widehat d_{in}^{\,}
+\sum_{n\in 1}\sum_{m \in 2}\mathcal{V}_{12}^{\,}\widehat d_{2m}^\dg \widehat d_{1n}^{\,}
\nonumber\\
&&
+\sum_{\alpha =L,R}\sum_{k \in \alpha}\sum_{i=1,2}\sum_{n \in i} V_{i\alpha}^{\,}\widehat c_{\alpha k}^\dg \widehat d_{in}^{\,} +h.c.~,
\end{eqnarray}
where $\varepsilon_{\alpha k}$ is the energy of $k$-momentum state in reservoir $\alpha$.  Each dot $i$ contains $N$ discrete energy levels denoted as $\varepsilon_{in}$, with $n\in[1,N]$,  which are assumed in the following to be regularly spaced by an energy difference equal to $\varepsilon_{0}$, thus one has $\varepsilon_{in}=\varepsilon_i+n\varepsilon_{0}$ with $\varepsilon_i$ the energy of the lowest level in dot $i$, equal to either~$\varepsilon_1$ or~$\varepsilon_2$. The notation $h.c.$ corresponds to the hermitian conjugate terms associated with the third and fourth contributions in Eq.~(\ref{H}). One assumes that the inter-dot coupling $\mathcal{V}_{12}$ between the states $|1n\rangle$ and $|2m\rangle$ in the dots, and the hopping integral $V_{i\alpha}$ between the states $|in\rangle$ in the dot $i$ and $|\alpha k\rangle$ in the reservoir~$\alpha$ depend neither on the indices $n$ and $m$, nor on the momentum~$k$.

The retarded Green function in the double quantum dot is a $2\times 2$ matrix given by\cite{Lavagna2020}
\begin{eqnarray}\label{FG_maintext}
&&\doubleunderline{{\bf G}^r}(\varepsilon) =\frac{1}{D^{r}(\varepsilon)}\nonumber\\
&&\times\left(
\begin{array}{cc}
\widetilde{\bf g}_1^{r}(\varepsilon) & \widetilde{\bf g}_1^{r}(\varepsilon)\mathbb{\widetilde \Sigma}_{12}^{r}(\varepsilon)\widetilde{\bf g}_2^{r}(\varepsilon)\\
\widetilde{\bf g}_2^{r}(\varepsilon)\mathbb{\widetilde\Sigma}_{21}^{r}(\varepsilon)\widetilde{\bf g}_1^{r}(\varepsilon) & \widetilde{\bf g}_2^{r}(\varepsilon)
\end{array}
\right)~,
\end{eqnarray}
where $D^{r}(\varepsilon) = 1-\widetilde{\bf g}_{1}^{r}(\varepsilon)\mathbb{\widetilde\Sigma}_{12}^{r}(\varepsilon)\widetilde{\bf g}_{2}^{r}(\varepsilon)\mathbb{\widetilde\Sigma}_{21}^{r}(\varepsilon)$ and  ${\widetilde{\bf g}_{i}^{r}}(\varepsilon) ={\bf g}_{i}^{r} (\varepsilon)/(1 - {{\mathbb{\widetilde\Sigma}}_{ii}^{r}}(\varepsilon){\bf g}_{i}^{r}(\varepsilon))$. In these expressions appear the retarded Green function of the disconnected dot~$i$, defined as ${\bf g}_i^{r}(\varepsilon)=\sum_{n\in i}g_{in}^{r}(\varepsilon)$ with $g_{in}^{r}(\varepsilon) = 1/(\varepsilon - \varepsilon_{in} + i0^+)$, and the total self-energy: $ \doubleunderline{\mathbb{\widetilde\Sigma}^{r}}(\varepsilon)= \sum_{\alpha=L,R}\doubleunderline{\mathbb{\Sigma}^r_\alpha}(\varepsilon)+\doubleunderline{\mathbb{\Sigma}^r_\mathrm{inter}}$, where the matrix $\doubleunderline{\mathbb{\Sigma}^{r}_\mathrm{inter}}$ is given by
\begin{eqnarray}
\doubleunderline{\mathbb{\Sigma}^{r}_\mathrm{inter}} =\left(
\begin{array}{cc}
0 & \mathcal{V}^*_{12}\\
\mathcal{V}^*_{21} & 0
\end{array}
\right)~.
\end{eqnarray}
The advanced Green function matrix $\doubleunderline{{\bf G}^a}(\varepsilon)$ is obtained from the retarded one by replacing the superscript $r$ by the superscript $a$ in the expression of $\doubleunderline{{\bf G}^r}(\varepsilon)$. Moreover, one has
$\doubleunderline{{\bf G}^{\lessgtr}}(\varepsilon)= \sum_{\alpha=L,R}\doubleunderline{{\bf G}^r}(\varepsilon)\doubleunderline{\mathbb{\Sigma}^\lessgtr_\alpha}(\varepsilon)\doubleunderline{{\bf G}^{a}}(\varepsilon)$. The noise is calculated from Eqs.~(\ref{SLL}) and (\ref{SLR}) in which the Green function given by Eq.~(\ref{FG_maintext}) is injected.

 In the following we consider only symmetrical dot-reservoir couplings, meaning that $\Gamma\equiv\Gamma_{L,11}=\Gamma_{R,22}$ and $\Gamma_{L,22}=\Gamma_{R,11}=0$ for a serial double dot, and $\Gamma\equiv\Gamma_{\alpha,ij}$, $\forall\{\alpha,i,j\}$, for a parallel double dot.

%%%%%%%%%%%%%%%%%%%%%%%%%%%%%%%%%%%%%%%%%%%%%%%%%%%%%%%%%%%%%%%%%%
%																 %
%																 %
%		DISCUSSION       											 %
%																 %
%																 %
%%%%%%%%%%%%%%%%%%%%%%%%%%%%%%%%%%%%%%%%%%%%%%%%%%%%%%%%%%%%%%%%%%

\begin{figure}[t]
\begin{center}
\includegraphics[width=4.cm]{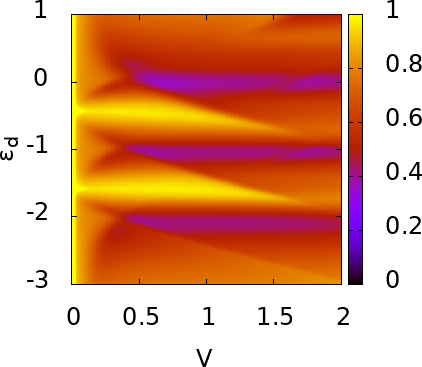}
\includegraphics[width=4.cm]{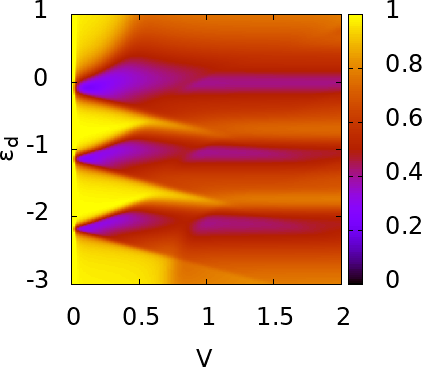}
\vspace*{-0.2cm}
\caption{Color-scale plot of the Fano factor $\mathcal{F}_{L}$ as a function of the bias voltage~$V$ and the detuning energy $\varepsilon_d$ for a double quantum dot in series, with three energy levels in each dot, equal to $\varepsilon_{in}=\varepsilon_i+n\varepsilon_0$ where $n=1,2,3$, with (left)~$\varepsilon_1=0$, and  (right)~$\varepsilon_1=0.3$. The other parameters are $\varepsilon_0=1$, $k_BT_{L,R}=0.01$, $\Gamma=0.1$, and $\mathcal{V}_{12}=0.2$.}\label{figure_FanoFLL_N1}
\end{center}
\end{figure}

\begin{figure}[t]
\begin{center}
\includegraphics[width=4.cm]{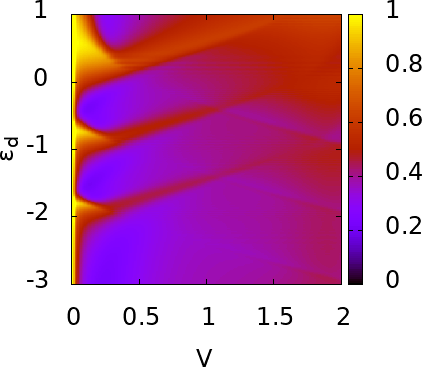}
\includegraphics[width=4.cm]{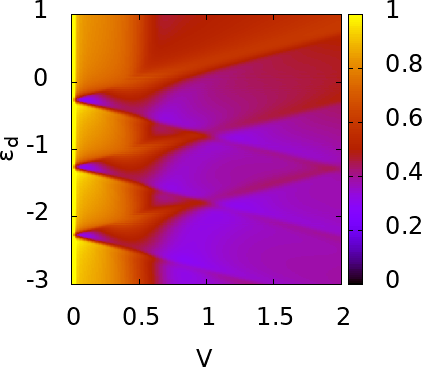}
\vspace*{-0.2cm}
\caption{Same as in Fig.~\ref{figure_FanoFLL_N1} for a double quantum dot in parallel.}\label{figure_FanoFLL_N2}
\end{center}
\end{figure}

\subsection{Fano factor}

We first study the Fano factor, defined as $\mathcal{F}_{\alpha}=\mathcal{S}_{\alpha\alpha}(0)/(e\langle \widehat I_\alpha\rangle)$, in order to identify the required conditions for obtaining a reduction of the zero-frequency noise compared to the current, i.e., such that one has $\mathcal{F}_{\alpha}\ll~1$. A priori, such a reduction should occur when the transmission through the double quantum dot is high, i.e. when the system leaves the Poissonian limit (for which $\mathcal{F}_{L}\approx 1$). This regime is hardly possible to study with a perturbative approach, but can be described within our approach which takes dot-reservoir and dot-dot couplings into account to all orders. Figure~\ref{figure_FanoFLL_N1} displays the color-scale plot of the Fano factor in the $L$ reservoir,~$\mathcal{F}_{\alpha}$, as a function of bias voltage~$eV=\mu_L-\mu_R$ and detuning energy $\varepsilon_d=\varepsilon_2-\varepsilon_1$, for a double quantum dot in series with $N=3$ energy levels in each dot ($n=1,2,3$). It shows that when the energy $\varepsilon_1$ of dot~1 is aligned with $\mu_L+\mu_R$, here equal to zero since we set $\mu_L=eV/2$ and $\mu_R=-eV/2$, the value of $\mathcal{F}_{L}$  at low voltage is usually 0.5 or higher, excepted in some  narrow purple stripes along the lines $\varepsilon_d=m\varepsilon_0$, with $m$ integer, when $eV\gtrsim 0.5$ (see left panel of Fig.~\ref{figure_FanoFLL_N1}). On the contrary, when the energy~$\varepsilon_1$ of dot~1 is not aligned with $\mu_L+\mu_R$ (the most likely situation), one observes that the value of $\mathcal{F}_{L}$ is strongly reduced at low voltage for some values of the detuning energy (see the purple color areas in the right panel of Fig.~\ref{figure_FanoFLL_N1}), opening up the possibility of reducing noise, in comparison to the current, even at low voltage. To explain the values taken by the Fano factor $\mathcal{F}_{L}$, one has to consider and compare the values of the current $\langle \widehat I_L\rangle$ and noise $\mathcal{S}_{LL}(0)$. For $\varepsilon_1=0$ and low voltage ($eV\lesssim 0.5$), the current and the transmission are weak so that the system is in the Poissonian limit, with  $\mathcal{F}_{L}\approx 1$, as observed in the left panel of Fig.~\ref{figure_FanoFLL_N1}. As soon as $\varepsilon_1\ne 0$, the transmission through the left reservoir increases and the system leaves the  Poissonian limit, leading to a reduction of $\mathcal{F}_{L}$ at low voltage when the dot energy levels align, as observed in the right panel of Fig.~\ref{figure_FanoFLL_N1}. In the case of a double quantum dot arranged in parallel, the reduction of the Fano factor is more pronounced. Indeed, when the detuning energy is aligned with $\mu_L+\mu_R$ one observes a strong reduction of noise at low voltage (see the purple color areas in the left panel of Fig.~\ref{figure_FanoFLL_N2}). Even for $\varepsilon_d\ne \mu_L+\mu_R$, there are large purple regions where $\mathcal{F}_{L}\approx 0.5$ is reduced (see the right panel of Fig.~\ref{figure_FanoFLL_N2}). This is due to the fact that for a double quantum dot in parallel, there exists two transmission paths, either through dot 1 or dot 2, leading to an increasing of the total transmission and by consequence to a reduction of the Fano factor. These results help us to identify the areas in the $(V,\varepsilon_d)$-plane where the noise is reduced.

\begin{figure}[t]
\begin{center}
\includegraphics[width=4.cm]{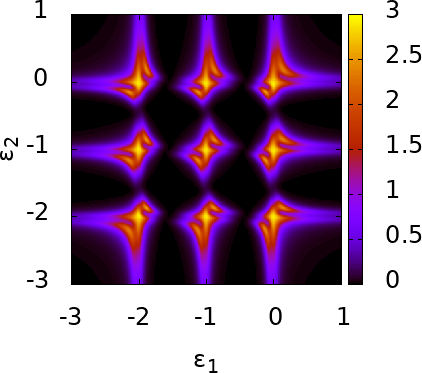}
\includegraphics[width=4.cm]{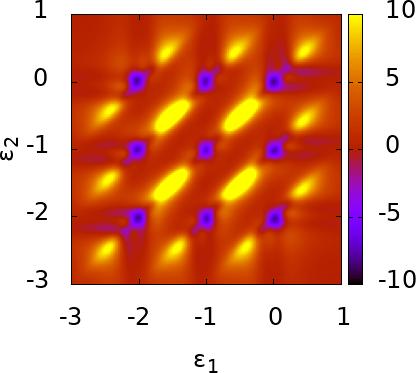}
\vspace*{-0.2cm}
\caption{Color-scale plot of the $\Delta T$-noise as a function of~$\varepsilon_1$ and $\varepsilon_2$ for a double quantum dot in series (a.u.), with three energy levels in each dot, equal to $\varepsilon_{in}=\varepsilon_i+n\varepsilon_0$ where $n=1,2,3$, with (left)~$k_BT=0.01$, $\Gamma=0.1$, $\mathcal{V}_{12}=0.1$, and (right) $k_BT=0.1$, $\Gamma=0.05$, $\mathcal{V}_{12}=0.1$. The other parameters are $\delta T=T/2$, $\varepsilon_0=1$, and $V=0$.}\label{figure_DeltaTnoise_N1}
\end{center}
\end{figure}

\begin{figure}[t]
\begin{center}
\includegraphics[width=4.cm]{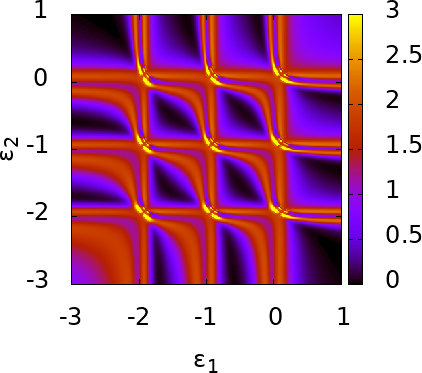}
\includegraphics[width=4.cm]{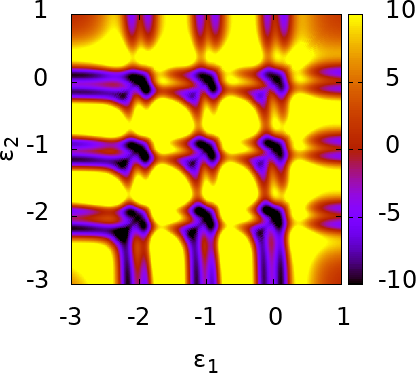}
\vspace*{-0.2cm}
\caption{Same as in Fig.~\ref{figure_DeltaTnoise_N1} for a double quantum dot in parallel.}\label{figure_DeltaTnoise_N2}
\end{center}
\end{figure}

\subsection{$\Delta T$-noise in the $(\varepsilon_1,\varepsilon_2)$-plane}

\begin{figure}[t]
\begin{center}
\includegraphics[width=8cm]{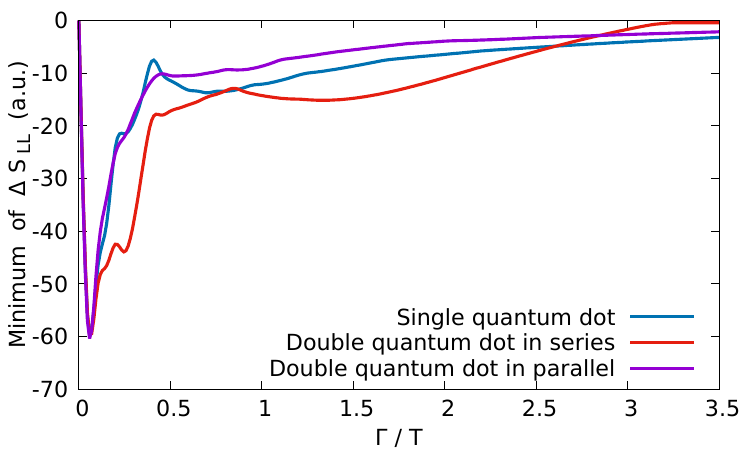}
\caption{Value of the minimum of the $\Delta T$-noise $\Delta \mathcal{S}_{LL}$, according to $\varepsilon_1$ and  $\varepsilon_2$, as a function of the ratio $\Gamma/T$ at $V=0$, $\mathcal{V}_{12}=0.1$, $\Gamma=0.1$, and $\varepsilon_0=1$. One takes $\delta T=T/2$. For both simple and double quantum dots, three energy levels in each dot have been included. In the parallel geometry case, a scaling factor 1/2 has been added in order to be able to compare the different curves.}\label{figure_DeltaTnoise_N3}
\end{center}
\end{figure}

Next, we turn our interest to the $\Delta T$-noise defined as
$\Delta \mathcal{S}_{\alpha\alpha}=\mathcal{S}_{\alpha\alpha}^{\delta T}(0)-\mathcal{S}_{\alpha\alpha}^0(0)$,
where $\mathcal{S}_{\alpha\alpha}^{\delta T}(0)$ is the zero-frequency auto-correlator at zero voltage when the reservoir are cooled down to distinct temperatures, i.e., $T_L=T+\delta T/2$ and $T_R=T-\delta T/2$, and $\mathcal{S}_{\alpha\alpha}^{0}(0)$ is the zero-frequency auto-correlator  at zero-voltage when the two reservoirs are at the same temperatures, i.e., $T_{L,R}=T$. Figure~\ref{figure_DeltaTnoise_N1} displays the color-scale plots of the $\Delta T$-noise in a double quantum dot in series for two different sets of parameters. It shows that in the regime where $\Gamma\gtrsim T$, the $\Delta T$-noise remains positive (see left panel of Fig.~\ref{figure_DeltaTnoise_N1}), whereas in the regime where $\Gamma\lesssim T$, $\Delta T$-noise changes sign (see right panel of Fig.~\ref{figure_DeltaTnoise_N1}), meaning that the noise is reduced in some areas of the $(\varepsilon_1,\varepsilon_2)$-plane when a temperature gradient between the two reservoirs is applied. The fact that a change of behavior is observed in the $\Delta T$-noise, when reducing~$\Gamma$, is in agreement with what has been obtained and explained in Ref.~\cite{Zhang2022}, with a change of behavior at $\Gamma/T\approx 2.6$, mainly due to the fact that the energy dependence of the transmission amplitude~$\doubleunderline{t_{\alpha\alpha}}(\varepsilon)$ and transmission coefficients~$\doubleunderline{\mathcal{T}_{\alpha\alpha}}(\varepsilon)$ become relevant when $\Gamma\lesssim T$. In the case of a double quantum dot in parallel, such a change in $\Delta T$-noise behavior is also observed between the two regimes $\Gamma\lesssim T$ and $\Gamma\gtrsim T$ (see Fig.~\ref{figure_DeltaTnoise_N2}). In order to further understand this phenomenon, we have reported in Fig.~\ref{figure_DeltaTnoise_N3} the evolution of the $\Delta T$-noise minimum as a function of $\Gamma/T$ for three different quantum dot geometries: single dot, double dot in series, and double dot in parallel. We see that the value of this minimum is negative at low $\Gamma/T$ and converges to zero at $\Gamma/T\approx 3$ whatever the geometry is. Again, this variation is in agreement with previous results obtained for $\Delta T$-noise in quantum systems characterized by energy-dependent transmission\cite{Zhang2022}. This is an important result insofar as it means that there is an operating regime in which noise can be reduced by applying a temperature gradient between the right and left reservoirs, in both single and double quantum dot systems.

\begin{figure}[t]
\begin{center}
\includegraphics[height=3.2cm]{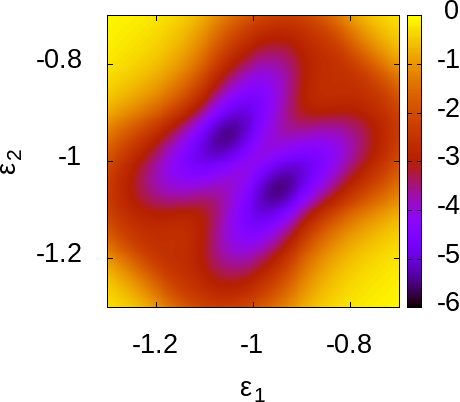}
\includegraphics[height=3.2cm]{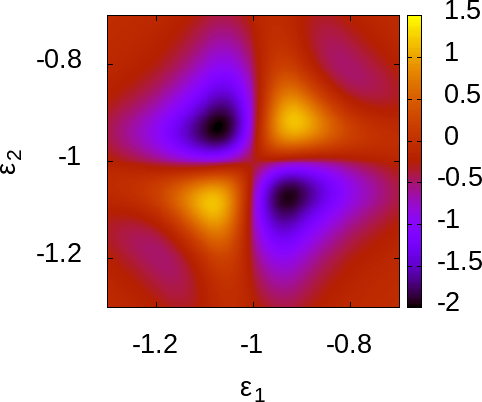}
\vspace*{-0.2cm}
\caption{Color-scale plot of the real part of the cross-correlator $\mathcal{S}_{LR}(\omega)$ as a function of~$\varepsilon_1$ and $\varepsilon_2$ for a double quantum dot in series (a.u.), with three energy levels in each dot, equal to $\varepsilon_{in}=\varepsilon_i+n\varepsilon_0$ where $n=1,2,3$, with (left)~$\hbar\omega=0$, and (right) $\hbar\omega=0.1$. The other parameters are $eV=0.2$, $\Gamma=0.1$, $\mathcal{V}_{12}=0.1$, $k_BT_{L,R}=0.01$, and $\varepsilon_0=1$.}\label{figure_Cross}
\end{center}
\end{figure}

\begin{figure}[t]
\begin{center}
\includegraphics[height=3.2cm]{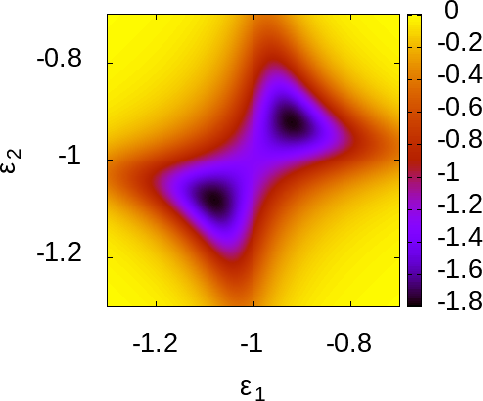}
\includegraphics[height=3.2cm]{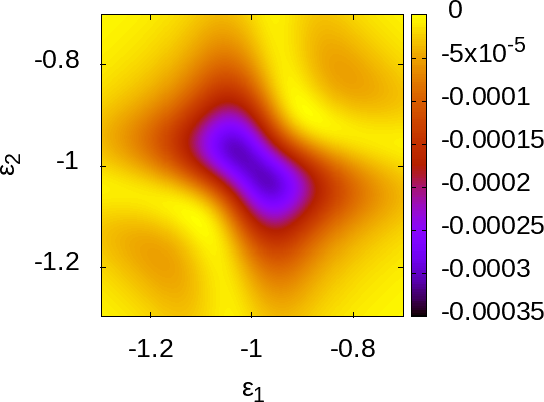}
\vspace*{-0.2cm}
\caption{Same as in Fig. \ref{figure_Cross} at $V=0$.}\label{figure_Cross_bis}
\end{center}
\end{figure}

\subsection{Cross-correlator in the $(\varepsilon_1,\varepsilon_2)$-plane}\label{ni_cc}

We now study the current-current cross-correlator for a double quantum dot coupled in series. Figure~\ref{figure_Cross} shows the real part of $\mathcal{S}_{LR}(\omega)$ near the honeycomb vertex located in the central region of the $(\varepsilon_1,\varepsilon_2)$-plane, at both zero-frequency ($\hbar\omega=0$) and finite-frequency ($\hbar\omega=0.1$). At zero-frequency, the current cross-correlator has a negative sign (see left panel of Fig.~\ref{figure_Cross}) as expected since $\mathcal{S}_{LR}(0)=-\mathcal{S}_{LL}(0)$ with positive current auto-correlator sign, whereas at finite-frequency the sign of the real part of the cross-correlator becomes positive in some areas of the $(\varepsilon_1,\varepsilon_2)$-plane (see right panel of Fig.~\ref{figure_Cross}). It deserves to be emphasized that the entire evolution of the current cross-correlator that we obtained at finite-frequency is in agreement with the experimental results presented in Ref.~\cite{McClure2007} which show a sign change in the vicinity of the honeycomb vertices. In these experiment, two capacitively coupled quantum dots are studied in the Coulomb blockade regime and the change of sign is attributed to Coulomb interactions. Generally, such a sign change in current cross-correlator is observed in interacting systems for which either a bosonic statistics, with bunching processes giving positive cross-correlator, or a fermionic statistics, with anti-bunching processes giving negative cross-correlator, are awaited\cite{Creux2006}. However, since we have not included yet Coulomb interactions in our model, this means that the explanation for the change in sign of the real part of the cross-correlator is here related to the fact that the frequency is taken non-zero. And identically to what is experimentally observed, one obtains a vanishing small finite-frequency cross-correlator at zero voltage,  as displayed in Fig.~\ref{figure_Cross_bis}, due to the fact that at low temperature the system can not emit noise at frequency larger than the bias voltage\cite{Aguado2000,Deblock2003,Fevrier2018}, so that at $\hbar\omega>eV$, one has $\mathcal{S}_{LR}(\omega)\approx\mathcal{S}_{LL}(\omega)\approx 0$.

%%%%%%%%%%%%%%%%%%%%%%%%%%%%%%%%%%%%%%%%%%%%%%%%%%%%%%%%%%%%%%%%%%
%																 %
%																 %
%		INTERACTING DQD											     %
%																 %
%																 %
%%%%%%%%%%%%%%%%%%%%%%%%%%%%%%%%%%%%%%%%%%%%%%%%%%%%%%%%%%%%%%%%%%

\section{Interacting double quantum dot}\label{interacting}

\subsection{Model}

In the presence of Coulomb interactions in the dots, the spin degrees of freedom and an additional contribution to the Hamiltonian have to be included, giving
\begin{eqnarray}\label{Hint}
\mathcal{\widehat H}&=&\sum_{\alpha =L,R}\sum_{k \in \alpha}\sum_{\sigma=\uparrow,\downarrow} \varepsilon_{\alpha k}^{\,}\widehat c_{\alpha k\sigma}^\dg \widehat c_{\alpha k\sigma}^{\,} \nonumber\\
&&
+\sum_{i=1,2}\sum_{n \in i}\sum_{\sigma=\uparrow,\downarrow} \varepsilon_{in}^{\,}\widehat d_{in\sigma}^\dg \widehat d_{in\sigma}^{\,}\nonumber\\
&&
+\sum_{n\in 1}\sum_{m \in 2}\sum_{\sigma=\uparrow,\downarrow}\mathcal{V}_{12}^{\,}\widehat d_{2m\sigma}^\dg \widehat d_{1n\sigma}^{\,}
\nonumber\\
&&
+\sum_{\alpha =L,R}\sum_{k \in \alpha}\sum_{i=1,2}\sum_{n \in i}\sum_{\sigma=\uparrow,\downarrow} V_{i\alpha}^{\,}\widehat c_{\alpha k\sigma}^\dg \widehat d_{in\sigma}^{\,} +h.c.\nonumber\\
&&
+U\sum_{i=1,2}\sum_{n \in i}\widehat d_{in\uparrow}^\dg\widehat d_{in\uparrow}\widehat d_{in\downarrow}^\dg\widehat d_{in\downarrow}~,
\end{eqnarray}
where $U$ is the Coulomb interaction strength. The Green function ${\widetilde{\bf g}_{i}^{r,a}}(\varepsilon)$ associated with each individual interacting dot $i$ is numerically calculated by using a self-consistent approach based on the resolution of the equations of motion governing the behavior of the dot Green functions in the presence of Coulomb interactions\cite{Roermund2010,Lavagna2015}. Once ${\widetilde{\bf g}_{i}^{r}}(\varepsilon)$ is obtained, we insert its value in Eq.~(\ref{FG_maintext}) in order to obtain the Green function~$\doubleunderline{{\bf G}^r}(\varepsilon)$ of the double quantum dot. Finally, we insert $\doubleunderline{{\bf G}^r}(\varepsilon)$ in Eqs.~(\ref{SLL}) and (\ref{SLR}) to calculate the finite-frequency auto-correlator and cross-correlator in the presence of Coulomb interactions. The numerical results are presented hereinafter. For simplicity, we restrict our study to the case where each dot contains a single energy level, $\varepsilon_1$ and $\varepsilon_2$.

\begin{figure}[t]
\begin{center}
\includegraphics[height=3.7cm]{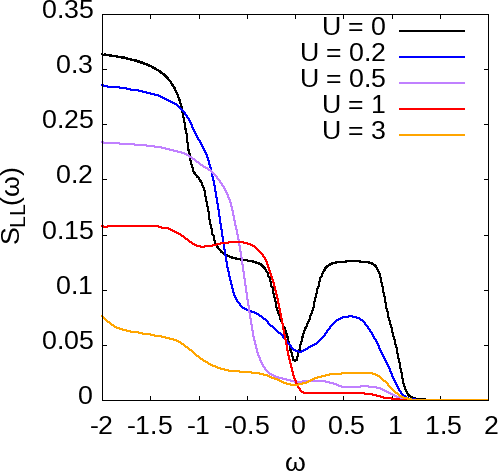}
\includegraphics[height=3.7cm]{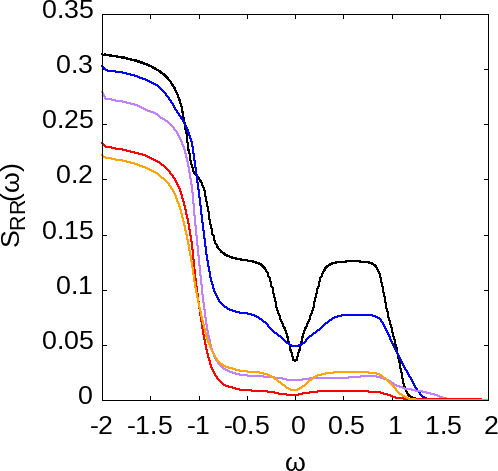}
\vspace*{-0.2cm}
\caption{Auto-correlators $\mathcal{S}_{LL}(\omega)$ and $\mathcal{S}_{RR}(\omega)$ as a function of~$\omega$ for a double quantum dot in series (a.u.) with a single energy level in each dot, equal to $\varepsilon_{1,2}=0$, with $eV=2$, $\Gamma=0.1$, $\mathcal{V}_{12}=0.1$, and $k_BT_{L,R}=0.01$.}\label{figure_FF_auto_series}
\end{center}
\end{figure}

\begin{figure}[t]
\begin{center}
\includegraphics[height=3.7cm]{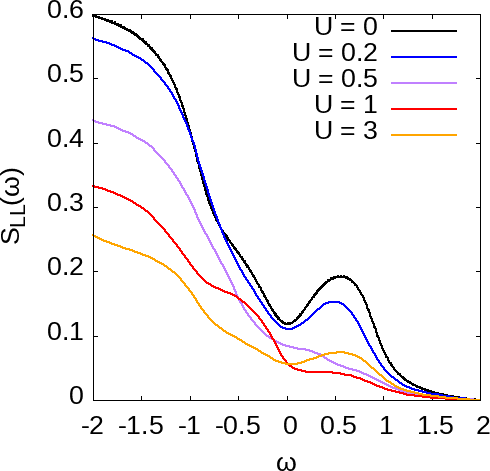}
\includegraphics[height=3.7cm]{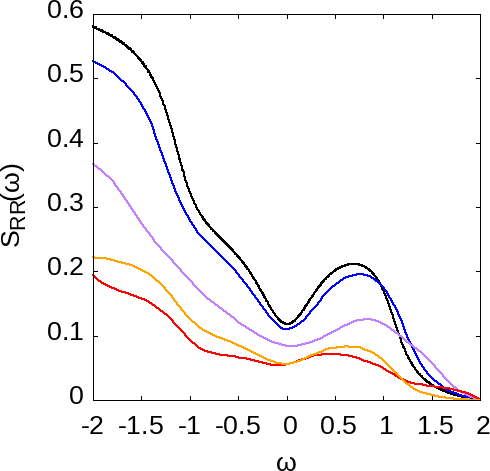}
\vspace*{-0.2cm}
\caption{Same as in Fig.~\ref{figure_FF_auto_series} for a double quantum dot arranged in parallel.}\label{figure_FF_auto_parallel}
\end{center}
\end{figure}

\subsection{Finite-frequency auto-correlator}

We start by studying the auto-correlators $\mathcal{S}_{LL}(\omega)$ and $\mathcal{S}_{RR}(\omega)$. Figures \ref{figure_FF_auto_series} and \ref{figure_FF_auto_parallel} show their evolution with the frequency $\omega$, at various values of Coulomb interaction~$U$, for a double quantum dot in series geometry and in parallel geometry respectively. We first remark that in the absence of interactions, i.e., for $U=0$, the auto-correlators in the left and right reservoirs coincide, $\mathcal{S}_{LL}(\omega)=\mathcal{S}_{RR}(\omega)$, due to the fact that one has taken a symmetrical potential profile through the double quantum dot, since one has $\mu_{L,R}=\pm eV/2$ and $\varepsilon_{1,2}=0$, similarly to what one has in single quantum dot\cite{Crepieux2018}. However, as soon as interactions are present, i.e., for $U\ne 0$, the auto-correlators in the left and right reservoirs differ, excepted at $\omega=0$ in the parallel case. The reason for this is that Coulomb interactions introduce a kind of asymmetry in the potential profile, leading to different left and right auto-correlators.

In Fig. \ref{figure_FF_auto_series}, we observe a staircase-like behavior with a smoothing of the auto-correlator amplitude in the presence of Coulomb interactions. Such a behavior has been already predicted using perturbative approach calculating symmetrized noise\cite{Shi2016}. However here we observe additional features such as those related to emission and absorption noises since we calculate non-symmetrized noise, with a cancellation of  $\mathcal{S}_{LL}(\omega)$ and $\mathcal{S}_{RR}(\omega)$ when the frequency becomes larger that the voltage. This allows to reveal a notable difference between what we observe for a double dot arranged in series (see Fig.~\ref{figure_FF_auto_series}) and in parallel (see Fig.~\ref{figure_FF_auto_parallel}): even in the presence of Coulomb interactions, the values of auto-correlators strongly reduce at $\hbar\omega>eV/2$ in the series case, whereas they remain non-zero in the parallel case, up to  $\hbar\omega\approx eV$. Indeed, in the former case, dot 1 is only connected to the left reservoir, whose chemical potential is equal to $\mu_L=eV/2$, whereas in the latter case, dot 1 is connected to both left and right reservoirs (see Fig.~\ref{figure0}) with voltage bias precisely equals to $eV=\mu_L-\mu_R$.

\begin{figure}[t]
\begin{center}
\includegraphics[height=3.7cm]{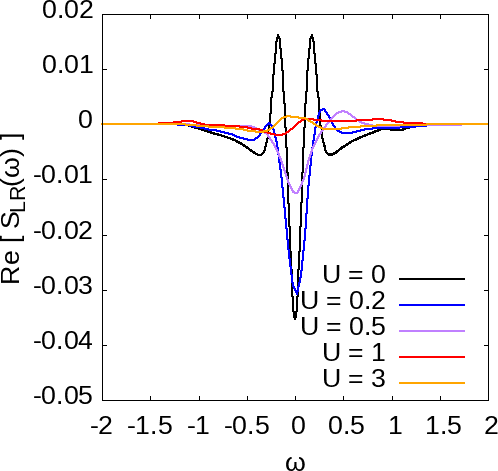}
\includegraphics[height=3.7cm]{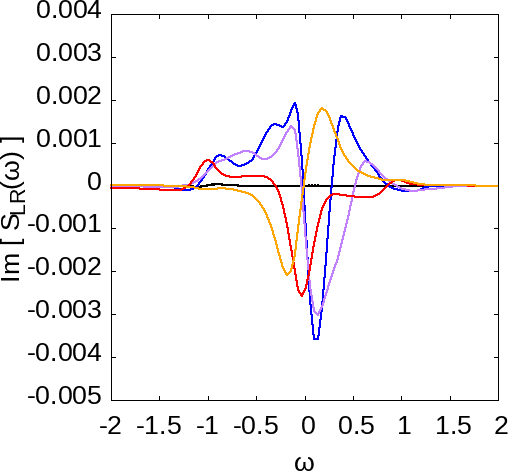}
\vspace*{-0.2cm}
\caption{Real and imaginary part of the cross-correlator $\mathcal{S}_{LR}(\omega)$ as a function of~$\omega$ for a double quantum dot in series~(a.u.) with a single energy level in each dot, equal to $\varepsilon_{1,2}=0$, with $eV=2$, $\Gamma=0.1$, $\mathcal{V}_{12}=0.1$, and $k_BT_{L,R}=0.01$.}\label{figure_FF_cross_series}
\end{center}
\end{figure}

\begin{figure}[t]
\begin{center}
\includegraphics[height=3.7cm]{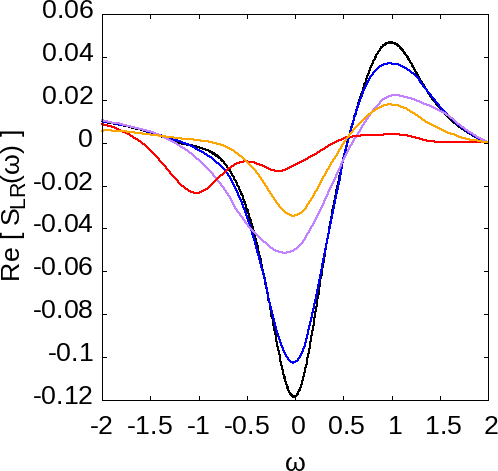}
\includegraphics[height=3.7cm]{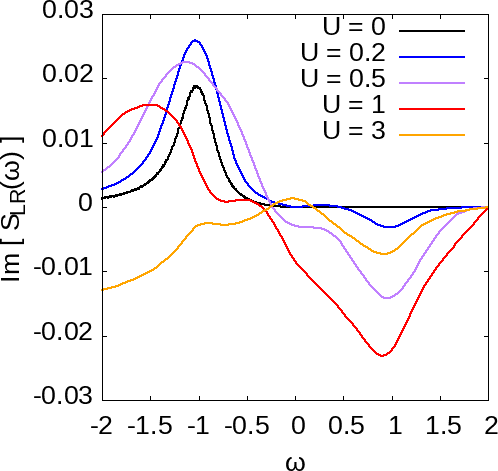}
\caption{Same as in Fig.~\ref{figure_FF_cross_series} for a double quantum dot arranged in parallel.}\label{figure_FF_cross_parallel}
\end{center}
\end{figure}

\subsection{Finite-frequency cross-correlator}

Finally, we consider the cross-correlators. Since one has $\mathcal{S}_{LR}(\omega)= \mathcal{S}^*_{RL}(\omega)$, we focus on $\mathcal{S}_{LR}(\omega)$. Figures \ref{figure_FF_cross_series} and \ref{figure_FF_cross_parallel} show the real and imaginary parts of the latter quantity as a function of frequency $\omega$, at increasing values of Coulomb interaction $U$, for both a double quantum dot in series and in parallel geometries. The most important result we obtain is that the sign of the real part of the cross-correlator can change under different circumstances. The first observation is that even in the absence of Coulomb interactions, the sign of $\mathrm{Re}\{\mathcal{S}_{LR}(\omega)\}$ varies with frequency (see black curves in Fig.~\ref{figure_FF_cross_series} and \ref{figure_FF_cross_parallel}): it is negative at zero frequency, then alternates between positive and negative signs as frequency increases. The second observation is that even at zero frequency, a change in the sign of $\mathrm{Re}\{\mathcal{S}_{LR}(\omega)\}$ occurs, in the case of a double quantum dot in series, when the interactions are sufficiently strong, here when $U=3$ (see orange curve in the left panel of Fig.~\ref{figure_FF_cross_series}). This result is in agreement with both experimental and theoretical studies\cite{Lambert2007,McClure2007}. Moreover, as in Ref.~\cite{Xu2022} which studies non-symmetrized noise, one observes for a double quantum dot in series that at $U=0$ the cross-correlator is symmetric in frequency, i.e., $\mathcal{S}_{LR}(-\omega)=\mathcal{S}_{LR}(\omega)$, whereas  at $U\ne 0$ it becomes asymmetrical, i.e., $\mathcal{S}_{LR}(-\omega)\ne\mathcal{S}_{LR}(\omega)$.  The explanation is the following: in the presence of Coulomb interactions, the spectral function $A(\varepsilon)$ is no longer an even function with energy since two asymmetrical peaks are pinned at chemical potentials $\mu_L$ and $\mu_R$ respectively\cite{Crepieux2018}, leading to an asymmetry in frequency. Finally, as it was the case for auto-correlators, and for the same reason, one obtains a vanishing $\mathcal{S}_{LR}(\omega)$ at $\hbar\omega>eV/2$ for a double quantum dot in series, whereas $\mathcal{S}_{LR}(\omega)$ vanishes when $\hbar\omega>eV$ for a double quantum dot in parallel. Our study has therefore shown that the change in sign of the real part of the cross-correlators can either result from Coulomb interactions, or from the fact that the measuring device operates at a non-zero frequency.

%%%%%%%%%%%%%%%%%%%%%%%%%%%%%%%%%%%%%%%%%%%%%%%%%%%%%%%%%%%%%%%%%%
%																 %
%																 %
%		CONCLUSION  											 %
%																 %
%																 %
%%%%%%%%%%%%%%%%%%%%%%%%%%%%%%%%%%%%%%%%%%%%%%%%%%%%%%%%%%%%%%%%%%

\section{Conclusion}\label{conclusion}

We have derived analytical expressions for the auto-correlators and cross-correlators of the current-current fluctuations in a double quantum dot that are valid at both zero and finite voltages, temperatures and frequencies, whatever the values of inter-dot and dot-reservoirs couplings are. We have highlighted specific features such as the reduction of the zero-frequency noise, as well as the possibility of having a negative $\Delta T$-noise, meaning that the noise can be further reduced by applying a temperature gradient between the two reservoirs. Moreover, it leads at finite frequency to a change of sign in the cross-correlator in the vicinity of honeycomb vertices. By including Coulomb interactions at the Hatree-Fock level, one shows that a change of sign in the cross-correlator can also be obtained due to the presence of interactions, even at zero-frequency. The approach presented in this work can be extended to take spin-orbit coupling into account, leading to effective g-factor\cite{Padawer2022} and Pauli spin blockade\cite{Lai2011,Lundberg2024}, which are all the more important to describe realistic situations in double quantum dot systems and spin qubits, as experimentally studied. We believe that the tools developed in this article will be invaluable in the realization of such project, which remains to be done.

%%%%%%%%%%%%%%%%%%%%%%%%%%%%%%%%%%%%%%%%%%%%%%%%%%%%%%%%%%%%%%%%%%
%																 %
%																 %
%		APPENDIX											     %
%																 %
%																 %
%%%%%%%%%%%%%%%%%%%%%%%%%%%%%%%%%%%%%%%%%%%%%%%%%%%%%%%%%%%%%%%%%%

\appendix

\section{Details of the noise calculation}\label{appendixA}

In the Appendices we present the calculation of finite-frequency noise in a double quantum dot performed in the flat wide-band limit using the non-equilibrium Green function technique. The main steps are the following: (i)~we perform a S-matrix expansion in order to perform the calculation at any order with coupling energies, (ii)~we use a Hartree-Fock level approximation in order to write two-particle Green functions as a product of two one-particle Green functions, and (iii)~we apply an analytical continuation (\ref{appendixA}); we next perform a Fourier transform (\ref{appendixB}) which allows to express the noise in terms of the Green function of the double quantum dot and of the self-energy of the reservoir (\ref{appendixC}); we finally insert transmission amplitudes and coefficients (\ref{appendixD}) to get the final expressions for both electrical current auto-correlators and cross-correlators (\ref{appendixE}).

\subsection{S-matrix expansion}

To express the Green functions $G_1^{cd,>}$, $G_2^{cd,>}$, $G_3^{cd,>}$ and $G_4^{cd,>}$ appearing in Eqs.~(\ref{G1dd}) to (\ref{G4dd}) in terms of products of two-particle Green function in the double quantum dot, $G^{dd}$, and Green function in the reservoir~$\alpha$, $g_{\alpha k}$, we perform a S-matrix expansion, where
\begin{eqnarray}
S=\sum_{j=0}^{\infty}\frac{(-i)^\ell}{\ell!}\int_C d\tau_1...\int_C d\tau_\ell \langle T_C \widetilde{H}_t(\tau_1)...\widetilde{H}_t(\tau_\ell) \rangle~,\nonumber\\
\end{eqnarray}
where $T_C$ is the time ordered operator on the contour~$C$, $\tau_1$ to $\tau_\ell$, the times on contour~$C$, and $\widetilde{H}_t$, the transfer Hamiltonian $H_t$ written in the interaction representation, where
\begin{eqnarray}
H_t= \sum_{\alpha =L,R}\sum_{k \in \alpha}\sum_{i=1,2}\sum_{n \in i} V_{i\alpha}^{\,}\widehat c_{\alpha k}^\dg \widehat d_{in}^{\,} +h.c.
\end{eqnarray}

After some transformations, the noise reads
\begin{eqnarray}\label{giniran00}
 && \mathcal{S}_{\alpha\beta}(\tau,\tau')=\frac{e^2}{\hbar^2} \Bigg( \sum_{k\in \alpha}\sum_{i,j=1,2}\sum_{n\in i,m\in j } V_{i\alpha}V_{j\alpha}^* \nonumber\\
 &&\times\Big[g_{\alpha k}(\tau',\tau)G_{in,jm}(\tau,\tau')+g_{\alpha k}(\tau,\tau')G_{in,jm}(\tau',\tau) \Big]\delta_{\alpha\beta} \nonumber \\
 &&+\sum_{k\in\alpha,k'\in \beta} \sum_{i,j=1,2}\sum_{p,q=1,2}\sum_{n\in i,m\in j }\sum_{n'\in p,m'\in q }
  \int_{-\infty}^\infty d\tau_1 \int_{-\infty}^\infty d\tau_2 \nonumber\\
&&  \times\Big[- V_{i\alpha}V_{j\beta} V_{p\alpha}^*V_{q\beta}^*g_{\alpha k}(\tau_1,\tau)g_{\beta k'}(\tau_2,\tau') G_1^{dd}(\tau,\tau',\tau_1,\tau_2) \nonumber \\
 &&+V_{i\alpha}V_{j\beta}^{\ast}V_{p\alpha}^*V_{q\beta}g_{\alpha k}(\tau_1,\tau)g_{\beta k'}(\tau',\tau_2) G_2^{dd}(\tau,\tau',\tau_1,\tau_2) \nonumber \\
&& -V_{i\alpha}^{\ast}V_{j\beta}V_{p\alpha}V_{q\beta}^* g_{\alpha k}(\tau,\tau_1)g_{\beta k'}(\tau_2,\tau') G_3^{dd}(\tau,\tau',\tau_1,\tau_2) \nonumber \\
 &&-V_{i\alpha}^{\ast}V_{j\beta}^{\ast} V_{p\alpha}V_{q\beta}g_{\alpha k}(\tau,\tau_1)g_{\beta k'}(\tau',\tau_2) G_4^{dd}(\tau,\tau',\tau_1,\tau_2) \Big] \Bigg)\nonumber\\
 &&-\langle \widehat I_\alpha \rangle\langle \widehat I_\beta \rangle~.
\end{eqnarray}
In this expression, $G_{in,jm}(\tau,\tau')=i \langle T_C \widehat d_{in}(\tau)\widehat d_{jm}^{\;\dag}(\tau') \rangle$ and $G_1^{dd}$, $G_2^{dd}$, $G_3^{dd}$, $G_4^{dd}$ are respectively the one-particle and the two-particle Green functions in the double quantum dot. The latter one are defined as\cite{HaugJauho2008}:
\begin{eqnarray}\label{G1dd}
G_1^{dd}(\tau,\tau',\tau_1,\tau_2)&=&i^2 \langle T_C \widehat d_{in}(\tau)\widehat d_{jm}(\tau')\widehat d_{pn'}^{\;\dag}(\tau_1)\widehat d_{qm'}^{\;\dag}(\tau_2) \rangle ~,\nonumber\\
\label{G2dd}
G_2^{dd}(\tau,\tau',\tau_1,\tau_2)&=&i^2 \langle T_C \widehat d_{in}(\tau)\widehat d_{jm}^{\;\dag}(\tau')\widehat d_{pn'}^{\;\dag}(\tau_1)\widehat d_{qm'}(\tau_2) \rangle ~,\nonumber\\
\label{G3dd}
G_3^{dd}(\tau,\tau',\tau_1,\tau_2)&=&i^2 \langle T_C \widehat d_{in}^{\;\dag}(\tau)\widehat d_{jm}(\tau')\widehat d_{pn'}(\tau_1)\widehat d_{qm'}^{\;\dag}(\tau_2) \rangle ~,\nonumber\\
\label{G4dd}
G_4^{dd}(\tau,\tau',\tau_1,\tau_2)&=&i^2 \langle T_C \widehat d_{in}^{\;\dag}(\tau)\widehat d_{jm}^{\;\dag}(\tau')\widehat d_{pn'}(\tau_1)\widehat d_{qm'}(\tau_2) \rangle ~.\nonumber
\end{eqnarray}

\subsection{Hartree-Fock approximation}

To evaluate the two-particle Green functions, we use the Hartree-Fock approximation. It gives
\begin{eqnarray}\label{G1dd}
G_1^{dd}(\tau,\tau',\tau_1,\tau_2)&=&G_{in,qm'}(\tau,\tau_2)G_{jm,pn'}(\tau',\tau_1)\nonumber\\
&-&G_{in,pn'}(\tau,\tau_1)G_{jm,qm'}(\tau',\tau_2) ~,\\
G_2^{dd}(\tau,\tau',\tau_1,\tau_2)&=&G_{in,jm}(\tau,\tau')G_{qm',pn'}(\tau_2,\tau_1)\nonumber\\
&-&G_{in,pn'}(\tau,\tau_1)G_{qm',jm}(\tau_2,\tau') ~,\\
G_3^{dd}(\tau,\tau',\tau_1,\tau_2)&=&-G_{jm,in}(\tau',\tau)G_{pn',qm'}(\tau_1,\tau_2) \nonumber\\
&+&G_{pn',in}(\tau_1,\tau)G_{jm,qm'}(\tau',\tau_2) ~,\\\label{G4dd}
G_4^{dd}(\tau,\tau',\tau_1,\tau_2)&=&G_{qm',in}(\tau_2,\tau)G_{pn',jm}(\tau_1,\tau')\nonumber\\
&-&G_{pn',in}(\tau_1,\tau)G_{qm',jm}(\tau_2,\tau')  ~.
\end{eqnarray}

By separating the contributions to the noise into connected and disconnected parts\cite{HaugJauho2008}: $ \mathcal{S}_{\alpha\beta}(\tau,\tau')= \mathcal{S}_{\alpha\beta}^\mathrm{dis}(\tau,\tau')+ \mathcal{S}_{\alpha\beta}^\mathrm{con}(\tau,\tau')$, one can show that $ \mathcal{S}_{\alpha\beta}^\mathrm{dis}(\tau,\tau')=\langle \widehat I_\alpha \rangle\langle \widehat I_\beta \rangle$, meaning that only the connected part contributes to the noise, i.e., the first contributions in the right-hand-side of Eqs.~(\ref{G1dd}) to (\ref{G4dd}). In consequence, one gets:
\begin{eqnarray}\label{hfa}
 \mathcal{S}_{\alpha\beta}(\tau,\tau')&=&\frac{e^2}{\hbar^2}\Bigg(  \sum_{k\in \alpha}\sum_{i,j=1,2}\sum_{n\in i,m\in j } V_{i\alpha}V_{j\alpha}^*\nonumber\\
&&\times\Big[g_{\alpha k}(\tau',\tau)G_{in,jm}(\tau,\tau')\nonumber\\
&&+g_{\alpha k}(\tau,\tau')G_{in,jm}(\tau',\tau) \Big]\delta_{\alpha\beta} \nonumber\\
 &&+\sum_{k\in\alpha,k'\in \beta} \sum_{i,j=1,2}\sum_{p,q=1,2}\nonumber\\
&&\times \sum_{n\in i,m\in j }\sum_{n'\in p,m'\in q }\int_{-\infty}^\infty d\tau_1 \int_{-\infty}^\infty d\tau_2\nonumber \\
 &&\times \Big[- V_{i\alpha}V_{j\beta} V_{p\alpha}^*V_{q\beta}^*g_{\alpha k}(\tau_1,\tau)g_{\beta k'}(\tau_2,\tau')\nonumber\\
&&\times G_{in,qm'}(\tau,\tau_2)G_{jm,pn'}(\tau',\tau_1)  \nonumber\\
 &&
 +V_{i\alpha}V_{j\beta}^{\ast}V_{p\alpha}^*V_{q\beta}g_{\alpha k}(\tau_2,\tau)g_{\beta k'}(\tau',\tau_1)\nonumber\\
&&\times G_{in,jm}(\tau,\tau')G_{qm',pn'}(\tau_2,\tau_1) \nonumber \\
 &&+V_{i\alpha}^{\ast}V_{j\beta}V_{p\alpha}V_{q\beta}^*g_{\alpha k}(\tau,\tau_1)g_{\beta k'}(\tau_2,\tau')\nonumber\\
&&\times G_{jm,in}(\tau',\tau)G_{pn',qm'}(\tau_1,\tau_2)  \nonumber\\
 &&
 -V_{i\alpha}^{\ast}V_{j\beta}^{\ast} V_{p\alpha}V_{q\beta}g_{\alpha k}(\tau,\tau_1)g_{\beta k'}(\tau',\tau_2)\nonumber\\
&&\times G_{qm',in}(\tau_2,\tau)G_{pn',jm}(\tau_1,\tau')\Big] \Bigg)~.
\end{eqnarray}

\subsection{Analytical continuation}\label{comp}

By applying an analytical continuation we find $ \mathcal{S}_{\alpha\beta}(t,t')= \sum_{\ell=1}^5\mathcal{P}_\ell(t,t')$, where
\begin{eqnarray}
\label{P1}
\mathcal{P}_1(t,t')&=&\frac{e^2}{\hbar^2}  \sum_{k\in \alpha}\sum_{i,j=1,2}\sum_{n\in i,m\in j } V_{i\alpha}V_{j\alpha}^*\nonumber\\
&&\times\Big[g_{\alpha k}^{<}(t',t)G_{in,jm}^>(t,t')\nonumber\\
&&+g_{\alpha k}^{>}(t,t')G_{in,jm}^<(t',t) \Big]\delta_{\alpha\beta}~,
\end{eqnarray}
\begin{eqnarray}
\label{P2}
 \mathcal{P}_2(t,t')&=&-\frac{e^2}{\hbar^2} \sum_{k\in\alpha,k'\in \beta} \sum_{i,j=1,2}\sum_{p,q=1,2}\nonumber\\
&&\times\sum_{n\in i,m\in j }\sum_{n'\in p,m'\in q } V_{i\alpha}V_{j\beta} V_{p\alpha}^*V_{q\beta}^* \nonumber\\
  \label{P2}
 &&\times \int_{-\infty}^\infty dt_1\Big[ G_{jm,pn'}^{r}(t',t_1)g_{\alpha k}^{<}(t_1,t)\nonumber\\
 &&+G_{jm,pn'}^{<}(t',t_1)g_{\alpha k}^{a}(t_1,t) \Big] \nonumber\\
 &&\times \int_{-\infty}^\infty dt_2 \Big[ G_{in,qm'}^{>}(t,t_2)g_{\beta k'}^{a}(t_2,t')\nonumber\\
 &&+G_{in,qm'}^{r}(t,t_2)g_{\beta k'}^{>}(t_2,t') \Big]~,
 \end{eqnarray}
 \begin{eqnarray}
 \label{P3}
&&\mathcal{P}_3(t,t')=\frac{e^2}{\hbar^2}\sum_{k\in\alpha,k'\in \beta} \sum_{i,j=1,2}\sum_{p,q=1,2}\nonumber\\
&&\times\sum_{n\in i,m\in j }\sum_{n'\in p,m'\in q } V_{i\alpha}V_{j\beta}^{\ast}V_{p\alpha}^*V_{q\beta} G_{in,jm}^>(t,t') \nonumber \\
 &&\times  \int_{-\infty}^\infty dt_1 \int_{-\infty}^\infty dt_2\nonumber\\
 &&\times\Big[ g_{\beta k'}^{r}(t',t_1)G_{qm',pn'}^{r}(t_2,t_1)g_{\alpha k}^{<}(t_2,t)\nonumber \\
 &&+g_{\beta k'}^{r}(t',t_1)G_{qm',pn'}^{<}(t_2,t_1)g_{\alpha k}^{a}(t_2,t)\nonumber\\
 &&+g_{\beta k'}^{<}(t',t_1)G_{qm',pn'}^{a}(t_2,t_1)g_k^{a}(t_2,t) \Big]~,
 \end{eqnarray}
 \begin{eqnarray}
 \label{P4}
&&\mathcal{P}_4(t,t')=\frac{e^2}{\hbar^2}\sum_{k\in\alpha,k'\in \beta} \sum_{i,j=1,2}\sum_{p,q=1,2}\nonumber\\
&&\times\sum_{n\in i,m\in j }\sum_{n'\in p,m'\in q } V_{i\alpha}^{\ast}V_{j\beta}V_{p\alpha}V_{q\beta}^* G_{jm,in}^<(t',t) \nonumber \\
 &&\times \int_{-\infty}^\infty dt_1 \int_{-\infty}^\infty dt_2\nonumber\\
 &&\times\Big[ g_{\alpha k}^{>}(t,t_1)G_{pn',qm'}^{a}(t_1,t_2)g_{\beta k'}^{a}(t_2,t')\nonumber \\
 &&+g_{\alpha k}^{r}(t,t_1)G_{pn',qm'}^{>}(t_1,t_2)g_{\beta k'}^{a}(t_2,t')\nonumber\\
 &&+g_{\alpha k}^{r}(t,t_1)G_{pn',qm'}^{r}(t_1,t_2)g_{\beta k'}^{>}(t_2,t') \Big]~,
 \end{eqnarray}
 and,
 \begin{eqnarray}
 \label{P5}
\mathcal{P}_5(t,t')&=&-\frac{e^2}{\hbar^2}\sum_{k\in\alpha,k'\in \beta} \sum_{i,j=1,2}\sum_{p,q=1,2}\nonumber\\
&&\times\sum_{n\in i,m\in j }\sum_{n'\in p,m'\in q } V_{i\alpha}^{\ast}V_{j\beta}^{\ast} V_{p\alpha}V_{q\beta}\nonumber\\
 &&\times \int_{-\infty}^\infty dt_1 \Big[ g_{\alpha k}^{>}(t,t_1)G_{pn',jm}^{a}(t_1,t')\nonumber\\
 &&+g_{\alpha k}^{r}(t,t_1)G_{pn',jm}^{>}(t_1,t') \Big] \nonumber\\
 &&\times\int_{-\infty}^\infty dt_2 \Big[ g_{\beta k'}^{r}(t',t_2)G_{qm',in}^{<}(t_2,t)\nonumber\\
 &&+g_{\beta k'}^{<}(t',t_2)G_{qm',in}^{a}(t_2,t) \Big]~.
\end{eqnarray}

%%%%%%%%%%%%%%%%%%%%%%%%%%%%%%%%%%%%%%%%%%%%%%%%%%%%%%%
%                                                     %
%                                                     %
%          FF NOISE                                   %
%                                                     %
%                                                     %
%%%%%%%%%%%%%%%%%%%%%%%%%%%%%%%%%%%%%%%%%%%%%%%%%%%%%%%

\section{Fourier transform}\label{appendixB}

Since one has $ \mathcal{S}_{\alpha\beta}(t,t')= \sum_{\ell=1}^5\mathcal{P}_\ell(t,t')$, it leads to
\begin{eqnarray}
&& \mathcal{S}_{\alpha\beta}(\omega)= \int_{-\infty}^\infty dt e^{-i\omega t} \mathcal{S}_{\alpha\beta}(t,0)=\sum_{\ell=1}^5\mathcal{P}_\ell(\omega)~,
\end{eqnarray}
where $\mathcal{P}_\ell(\omega)=\int_{-\infty}^\infty dt e^{-i\omega t}\mathcal{P}_\ell(t,0)$. In this Appendix, we calculate $\mathcal{P}_\ell(\omega)$ for $\ell\in[1,5]$.

\subsection{Calculation of $\mathcal{P}_1(\omega)$}\label{four}

From Eq.~(\ref{P1}), one gets
\begin{eqnarray}
&&\mathcal{P}_1(\omega)=\frac{1}{(2\pi)^2}\frac{e^2}{\hbar^2} \sum_{k\in \alpha}\sum_{i,j=1,2}\sum_{n\in i,m\in j } V_{i\alpha}V_{j\alpha}^* \int_{-\infty}^\infty dt e^{-i\omega t} \nonumber \\
&&\times\Big[\int_{-\infty}^\infty d\varepsilon_1 \int_{-\infty}^\infty d\varepsilon_2  e^{i\varepsilon_1 t}e^{-i\varepsilon_2 t}g_{\alpha k}^{<}(\varepsilon_1)G_{in,jm}^>(\varepsilon_2) \nonumber \\
&&+\int_{-\infty}^\infty d\varepsilon_1 \int_{-\infty}^\infty d\varepsilon_2 e^{i\varepsilon_1 t}e^{-i\varepsilon_2 t}g_{\alpha k}^{>}(\varepsilon_2)G_{in,jm}^<(\varepsilon_1) \Big]\delta_{\alpha\beta}~, \nonumber \\
\end{eqnarray}
since
\begin{eqnarray}
 G_{in,jm}^{a,r,\lessgtr}(t,t')&=&(2\pi)^{-1}\int_{-\infty}^\infty e^{-i\epsilon(t-t')}G_{in,jm}^{a,r,\lessgtr}(\varepsilon)d\epsilon~,\\
 g_{\alpha k}^{a,r,\lessgtr}(t,t')&=&(2\pi)^{-1}\int_{-\infty}^\infty e^{-i\epsilon(t-t')}g_{\alpha k}^{a,r,\lessgtr}(\varepsilon)d\epsilon~,
\end{eqnarray}
where, in order to save space, we disregard the $\hbar$ factor in the exponential from now up to the end of the Appendix Section. This $\hbar$ factor is however restored in the expression of the noise given in the main text. We collect all the exponential arguments:
\begin{eqnarray}
&&\mathcal{P}_1(\omega)=\frac{1}{(2\pi)^2}\frac{e^2}{\hbar^2} \sum_{k\in \alpha}\sum_{i,j=1,2}\sum_{n\in i,m\in j } V_{i\alpha}V_{j\alpha}^*  \nonumber \\
&&\times\int_{-\infty}^\infty dt\int_{-\infty}^\infty d\varepsilon_1 \int_{-\infty}^\infty d\varepsilon_2 e^{it(\varepsilon_1-\varepsilon_2-\omega)}\nonumber \\
&&\times\Big[ g_{\alpha k}^{<}(\varepsilon_1)G_{in,jm}^>(\varepsilon_2)+g_{\alpha k}^{>}(\varepsilon_2)G_{in,jm}^<(\varepsilon_1) \Big]\delta_{\alpha\beta}~.
\end{eqnarray}
By performing integration on $t$, one obtains a Dirac delta function $2\pi \hbar \delta(\varepsilon_1-\varepsilon_2-\omega)$. Thus
\begin{eqnarray}\label{TFP1}
&&\mathcal{P}_1(\omega)= \frac{e^2}{h}\sum_{k\in \alpha}\sum_{i,j=1,2}\sum_{n\in i,m\in j }V_{i\alpha}V_{j\alpha}^* \int_{-\infty}^\infty d\varepsilon \nonumber \\
&&\times\Big[ g_{\alpha k}^{<}(\varepsilon)G_{in,jm}^>(\varepsilon-\omega)+g_{\alpha k}^{>}(\varepsilon-\omega)G_{in,jm}^<(\varepsilon) \Big]\delta_{\alpha\beta}~.\nonumber \\
\end{eqnarray}

\subsection{Calculation of $\mathcal{P}_2(\omega)$}

From Eq.~(\ref{P2}), one gets
\begin{eqnarray}
\mathcal{P}_2(\omega)&=&
-\frac{1}{(2\pi)^4}\frac{e^2}{\hbar^2} \sum_{k\in\alpha,k'\in \beta} \sum_{i,j=1,2}\sum_{p,q=1,2}\nonumber \\
&&\times\sum_{n\in i,m\in j }\sum_{n'\in p,m'\in q } V_{i\alpha}V_{j\beta} V_{p\alpha}^*V_{q\beta}^* \nonumber\\
 &&\times\int_{-\infty}^\infty dte^{-i\omega t} \int_{-\infty}^\infty dt_1\nonumber \\
 &&\times\Big[ G_{jm,pn'}^{r}(0,t_1)g_{\alpha k}^{<}(t_1,t)\nonumber \\
 &&+G_{jm,pn'}^{<}(0,t_1)g_{\alpha k}^{a}(t_1,t) \Big] \nonumber\\
 &&\times \int_{-\infty}^\infty dt_2 \Big[ G_{in,qm'}^{>}(t,t_2)g_{\beta k'}^{a}(t_2,0)\nonumber \\
 &&+G_{in,qm'}^{r}(t,t_2)g_{\beta k'}^{>}(t_2,0) \Big]~.
\end{eqnarray}
It leads to
\begin{eqnarray}
&&\mathcal{P}_2(\omega)=
-\frac{1}{(2\pi)^4}\frac{e^2}{\hbar^2} \sum_{k\in\alpha,k'\in \beta} \sum_{i,j=1,2}\sum_{p,q=1,2}\nonumber \\
&&\times\sum_{n\in i,m\in j }\sum_{n'\in p,m'\in q } V_{i\alpha}V_{j\beta} V_{p\alpha}^*V_{q\beta}^*\int_{-\infty}^\infty dt \nonumber\\
 &&\times \int_{-\infty}^\infty dt_1 \int_{-\infty}^\infty d\varepsilon_1 \int_{-\infty}^\infty d\varepsilon_2
e^{-i\omega t} e^{i\varepsilon_1 t_1}e^{-i\varepsilon_2 (t_1-t)}\nonumber \\
&&\times \Big[ G_{jm,pn'}^{r}(\varepsilon_1)g_{\alpha k}^{<}(\varepsilon_2)+G_{jm,pn'}^{<}(\varepsilon_1)g_{\alpha k}^{a}(\varepsilon_2) \Big] \nonumber\\
 &&\times  \int_{-\infty}^\infty dt_2\int_{-\infty}^\infty d\varepsilon_3 \int_{-\infty}^\infty d\varepsilon_4
  e^{-i\varepsilon_3(t-t_2)}e^{-i\varepsilon_4 t_2}\nonumber \\
&&\times \Big[ G_{in,qm'}^{>}(\varepsilon_3)g_{\beta k'}^{a}(\varepsilon_4)+G_{in,qm'}^{r}(\varepsilon_3)g_{\beta k'}^{>}(\varepsilon_4) \Big]~.
\end{eqnarray}
By performing integration on $t$, $t_1$ and $t_2$, one obtains the product of Dirac delta functions $(2\pi)^3 \hbar \delta(\varepsilon_1-\varepsilon_2)\delta(\varepsilon_3-\varepsilon_4)\delta(\varepsilon_2-\varepsilon_3-\omega)$. Thus
\begin{eqnarray}\label{TFP2}
\mathcal{P}_2(\omega)&=&
-\frac{e^2}{\hbar} \sum_{k\in\alpha,k'\in \beta} \sum_{i,j=1,2}\sum_{p,q=1,2}\nonumber \\
&&\times\sum_{n\in i,m\in j }\sum_{n'\in p,m'\in q } V_{i\alpha}V_{j\beta} V_{p\alpha}^*V_{q\beta}^* \nonumber\\
 &&\times\int_{-\infty}^\infty d\varepsilon
 \Big[ G_{jm,pn'}^{r}(\varepsilon)g_{\alpha k}^{<}(\varepsilon)\nonumber \\
 &&+G_{jm,pn'}^{<}(\varepsilon)g_{\alpha k}^{a}(\varepsilon) \Big] \nonumber\\
 &&\times
 \Big[ G_{in,qm'}^{>}(\varepsilon-\omega)g_{\beta k'}^{a}(\varepsilon-\omega)\nonumber \\
 &&+G_{in,qm'}^{r}(\varepsilon-\omega)g_{\beta k'}^{>}(\varepsilon-\omega) \Big]~.
\end{eqnarray}

\subsection{Calculation of $\mathcal{P}_3(\omega)$}

From Eq.~(\ref{P3}), one gets
\begin{eqnarray}
\mathcal{P}_3(\omega)&=&\frac{1}{(2\pi)^4}\frac{e^2}{\hbar^2}\sum_{k\in\alpha,k'\in \beta} \sum_{i,j=1,2}\sum_{p,q=1,2}\nonumber \\
&&\times\sum_{n\in i,m\in j }\sum_{n'\in p,m'\in q } V_{i\alpha}V_{j\beta}^{\ast}V_{p\alpha}^*V_{q\beta}\nonumber \\
 &&\times \int_{-\infty}^\infty dte^{-i\omega t}  G_{in,jm}^>(t,0)  \int_{-\infty}^\infty dt_1 \int_{-\infty}^\infty dt_2\nonumber \\
 &&\times\Big[ g_{\beta k'}^{r}(0,t_1)G_{qm',pn'}^{r}(t_2,t_1)g_{\alpha k}^{<}(t_2,t)\nonumber \\
 &&+g_{\beta k'}^{r}(0,t_1)G_{qm',pn'}^{<}(t_2,t_1)g_{\alpha k}^{a}(t_2,t)\nonumber \\
 &&+g_{\beta k'}^{<}(0,t_1)G_{qm',pn'}^{a}(t_2,t_1)g_k^{a}(t_2,t) \Big]~.
\end{eqnarray}
It leads to
\begin{eqnarray}
\mathcal{P}_3(\omega)&=&\frac{1}{(2\pi)^4}\frac{e^2}{\hbar^2}\sum_{k\in\alpha,k'\in \beta} \sum_{i,j=1,2}\sum_{p,q=1,2}\nonumber \\
&&\times\sum_{n\in i,m\in j }\sum_{n'\in p,m'\in q } V_{i\alpha}V_{j\beta}^{\ast}V_{p\alpha}^*V_{q\beta} \nonumber \\
 &&\times \int_{-\infty}^\infty dte^{-i\omega t}
 \int_{-\infty}^\infty d\varepsilon_1 \int_{-\infty}^\infty d\varepsilon_2 \nonumber \\
 &&\times\int_{-\infty}^\infty d\varepsilon_3 \int_{-\infty}^\infty d\varepsilon_4\int_{-\infty}^\infty dt_1 \int_{-\infty}^\infty dt_2 \nonumber \\
  &&\times e^{-i\varepsilon_1t} e^{i\varepsilon_2 t_2} e^{-i\varepsilon_3(t_2-t_1)}e^{-i\varepsilon_4(t_1-t)} G_{in,jm}^>(\varepsilon_1)\nonumber \\
  &&\times\Big[ g_{\beta k'}^{r}(\varepsilon_2)G_{qm',pn'}^{r}(\varepsilon_3)g_{\alpha k}^{<}(\varepsilon_4)\nonumber \\
 &&+g_{\beta k'}^{r}(\varepsilon_2)G_{qm',pn'}^{<}(\varepsilon_3)g_{\alpha k}^{a}(\varepsilon_4)\nonumber \\
 &&+g_{\beta k'}^{<}(\varepsilon_2)G_{qm',pn'}^{a}(\varepsilon_3)g_k^{a}(\varepsilon_4) \Big]~.
\end{eqnarray}
By performing integration on $t$, $t_1$ and $t_2$, one obtains the product of Dirac delta functions $(2\pi)^3 \hbar \delta(\varepsilon_2-\varepsilon_3)\delta(\varepsilon_3-\varepsilon_4)\delta(\varepsilon_4-\varepsilon_1-\omega)$. Thus
\begin{eqnarray}\label{TFP3}
&&\mathcal{P}_3(\omega)=\frac{e^2}{h}\sum_{k\in\alpha,k'\in \beta} \sum_{i,j=1,2}\sum_{p,q=1,2}\nonumber \\
&&\times\sum_{n\in i,m\in j }\sum_{n'\in p,m'\in q } V_{i\alpha}V_{j\beta}^{\ast}V_{p\alpha}^*V_{q\beta}  \nonumber \\
 &&\times
  \int_{-\infty}^\infty d\varepsilon G_{in,jm}^>(\varepsilon-\omega)\Big[ g_{\beta k'}^{r}(\varepsilon)G_{qm',pn'}^{r}(\varepsilon)g_{\alpha k}^{<}(\varepsilon)\nonumber \\
  &&+g_{\beta k'}^{r}(\varepsilon)G_{qm',pn'}^{<}(\varepsilon)g_{\alpha k}^{a}(\varepsilon)
  +g_{\beta k'}^{<}(\varepsilon)G_{qm',pn'}^{a}(\varepsilon)g_k^{a}(\varepsilon) \Big]~.\nonumber \\
\end{eqnarray}

\subsection{Calculation of $\mathcal{P}_4(\omega)$}

From Eq.~(\ref{P4}), one gets
\begin{eqnarray}
\mathcal{P}_4(\omega)&=&\frac{e^2}{\hbar^2}\sum_{k\in\alpha,k'\in \beta} \sum_{i,j=1,2}\sum_{p,q=1,2}\nonumber \\
&&\times\sum_{n\in i,m\in j }\sum_{n'\in p,m'\in q } V_{i\alpha}^{\ast}V_{j\beta}V_{p\alpha}V_{q\beta}^*  \nonumber \\
 &&\times \int_{-\infty}^\infty dte^{-i\omega t}G_{jm,in}^<(0,t) \int_{-\infty}^\infty dt_1 \int_{-\infty}^\infty dt_2 \nonumber \\
 &&\times \Big[ g_{\alpha k}^{>}(t,t_1)G_{pn',qm'}^{a}(t_1,t_2)g_{\beta k'}^{a}(t_2,0)\nonumber \\
 &&+g_{\alpha k}^{r}(t,t_1)G_{pn',qm'}^{>}(t_1,t_2)g_{\beta k'}^{a}(t_2,0)\nonumber \\
 &&+g_{\alpha k}^{r}(t,t_1)G_{pn',qm'}^{r}(t_1,t_2)g_{\beta k'}^{>}(t_2,0) \Big]~.
\end{eqnarray}
It leads to
\begin{eqnarray}
\mathcal{P}_4(\omega)&=&\frac{1}{(2\pi)^4}\frac{e^2}{\hbar^2}\sum_{k\in\alpha,k'\in \beta} \sum_{i,j=1,2}\sum_{p,q=1,2}\nonumber \\
&&\times\sum_{n\in i,m\in j }\sum_{n'\in p,m'\in q } V_{i\alpha}^{\ast}V_{j\beta}V_{p\alpha}V_{q\beta}^*  \nonumber \\
 &&\times \int_{-\infty}^\infty dte^{-i\omega t}\int_{-\infty}^\infty d\varepsilon_1 \int_{-\infty}^\infty d\varepsilon_2\nonumber \\
 &&\times\int_{-\infty}^\infty d\varepsilon_3 \int_{-\infty}^\infty d\varepsilon_4\int_{-\infty}^\infty dt_1 \int_{-\infty}^\infty dt_2 \nonumber \\
  &&\times e^{i\varepsilon_1t} e^{-i\varepsilon_2 (t-t_1)} e^{-i\varepsilon_3(t_1-t_2)}e^{-i\varepsilon_4t_2} \nonumber \\
  &&\times G_{jm,in}^<(\varepsilon_1) \Big[ g_{\alpha k}^{>}(\varepsilon_2)G_{pn',qm'}^{a}(\varepsilon_3)g_{\beta k'}^{a}(\varepsilon_4)\nonumber \\
 &&+g_{\alpha k}^{r}(\varepsilon_2)G_{pn',qm'}^{>}(\varepsilon_3)g_{\beta k'}^{a}(\varepsilon_4)\nonumber \\
 &&+g_{\alpha k}^{r}(\varepsilon_2)G_{pn',qm'}^{r}(\varepsilon_3)g_{\beta k'}^{>}(\varepsilon_4) \Big]~.
\end{eqnarray}
By performing integration on $t$, $t_1$ and $t_2$, one obtains the product of Dirac delta functions $(2\pi)^3 \hbar \delta(\varepsilon_2-\varepsilon_3)\delta(\varepsilon_3-\varepsilon_4)\delta(\varepsilon_1-\varepsilon_2-\omega)$. Thus
\begin{eqnarray}\label{TFP4}
&&\mathcal{P}_4(\omega)=\frac{e^2}{h}\sum_{k\in\alpha,k'\in \beta} \sum_{i,j=1,2}\sum_{p,q=1,2}\nonumber \\
&&\times\sum_{n\in i,m\in j }\sum_{n'\in p,m'\in q } V_{i\alpha}^{\ast}V_{j\beta}V_{p\alpha}V_{q\beta}^*
 \int_{-\infty}^\infty d\varepsilon G_{jm,in}^<(\varepsilon)\nonumber \\
 &&\times\Big[ g_{\alpha k}^{>}(\varepsilon-\omega)G_{pn',qm'}^{a}(\varepsilon-\omega)g_{\beta k'}^{a}(\varepsilon-\omega)\nonumber \\
 &&+g_{\alpha k}^{r}(\varepsilon-\omega)G_{pn',qm'}^{>}(\varepsilon-\omega)g_{\beta k'}^{a}(\varepsilon-\omega)\nonumber \\
 &&+g_{\alpha k}^{r}(\varepsilon-\omega)G_{pn',qm'}^{r}(\varepsilon-\omega)g_{\beta k'}^{>}(\varepsilon-\omega) \Big]~.
\end{eqnarray}

\subsection{Calculation of $\mathcal{P}_5(\omega)$}

From Eq.~(\ref{P5}), one gets
\begin{eqnarray}
&&\mathcal{P}_5(\omega)=-\frac{1}{(2\pi)^4}\frac{e^2}{\hbar^2}\sum_{k\in\alpha,k'\in \beta} \sum_{i,j=1,2}\sum_{p,q=1,2}\nonumber \\
&&\times\sum_{n\in i,m\in j }\sum_{n'\in p,m'\in q } V_{i\alpha}^{\ast}V_{j\beta}^{\ast} V_{p\alpha}V_{q\beta}\nonumber\\
 &&\times \int_{-\infty}^\infty dt e^{-i\omega t}\int_{-\infty}^\infty dt_1 \Big[ g_{\alpha k}^{>}(t,t_1)G_{pn',jm}^{a}(t_1,0)\nonumber \\
 &&+g_{\alpha k}^{r}(t,t_1)G_{pn',jm}^{>}(t_1,0) \Big] \nonumber\\
 &&\times\int_{-\infty}^\infty dt_2 \Big[ g_{\beta k'}^{r}(0,t_2)G_{qm',in}^{<}(t_2,t)\nonumber \\
 &&+g_{\beta k'}^{<}(0,t_2)G_{qm',in}^{a}(t_2,t) \Big]~.
\end{eqnarray}
It leads to
\begin{eqnarray}
&&\mathcal{P}_5(\omega)=-\frac{1}{(2\pi)^4}\frac{e^2}{\hbar^2}\sum_{k\in\alpha,k'\in \beta} \sum_{i,j=1,2}\sum_{p,q=1,2}\nonumber \\
&&\times\sum_{n\in i,m\in j }\sum_{n'\in p,m'\in q } V_{i\alpha}^{\ast}V_{j\beta}^{\ast} V_{p\alpha}V_{q\beta}\nonumber\\
 &&\times \int_{-\infty}^\infty  e^{-i\omega t} dt
 \int_{-\infty}^\infty d\varepsilon_1 \int_{-\infty}^\infty d\varepsilon_2 \nonumber\\
 &&\times\int_{-\infty}^\infty d\varepsilon_3 \int_{-\infty}^\infty d\varepsilon_4\int_{-\infty}^\infty dt_1 \int_{-\infty}^\infty dt_2 \nonumber \\
  &&\times e^{-i\varepsilon_1(t-t_1)} e^{-i\varepsilon_2 t_1} e^{i\varepsilon_3t_2}e^{-i\varepsilon_4(t_2-t)}\nonumber \\
&&\times \Big[ g_{\alpha k}^{>}(\varepsilon_1)G_{pn',jm}^{a}(\varepsilon_2)+g_{\alpha k}^{r}(\varepsilon_1)G_{pn',jm}^{>}(\varepsilon_2) \Big] \nonumber\\
 &&\times\Big[ g_{\beta k'}^{r}(\varepsilon_3)G_{qm',in}^{<}(\varepsilon_4)+g_{\beta k'}^{<}(\varepsilon_3)G_{qm',in}^{a}(\varepsilon_4) \Big]~.\nonumber \\
\end{eqnarray}
By performing integration on $t$, $t_1$ and $t_2$, one obtains the product of Dirac delta functions $(2\pi)^3 \hbar \delta(\varepsilon_2-\varepsilon_3)\delta(\varepsilon_3-\varepsilon_4)\delta(\varepsilon_1-\varepsilon_2-\omega)$. Thus
\begin{eqnarray}\label{TFP5}
\mathcal{P}_5(\omega)&=&-\frac{e^2}{h}\sum_{k\in\alpha,k'\in \beta} \sum_{i,j=1,2}\sum_{p,q=1,2}\nonumber \\
&&\times\sum_{n\in i,m\in j }\sum_{n'\in p,m'\in q } V_{i\alpha}^{\ast}V_{j\beta}^{\ast} V_{p\alpha}V_{q\beta}\nonumber\\
 &&\times
 \Big[ g_{\alpha k}^{>}(\varepsilon-\omega)G_{pn',jm}^{a}(\varepsilon-\omega)\nonumber \\
 &&+g_{\alpha k}^{r}(\varepsilon-\omega)G_{pn',jm}^{>}(\varepsilon-\omega) \Big] \nonumber\\
 &&\times\Big[ g_{\beta k'}^{r}(\varepsilon)G_{qm',in}^{<}(\varepsilon)+g_{\beta k'}^{<}(\varepsilon)G_{qm',in}^{a}(\varepsilon) \Big]~.\nonumber \\
\end{eqnarray}

%%%%%%%%%%%%%%%%%%%%%%%%%%%%%%%%%%%%%%%%%%%%%%%%%%%%%%%
%                                                     %
%                                                     %
%          GENERAL EXPRESSION                                    %
%                                                     %
%                                                     %
%%%%%%%%%%%%%%%%%%%%%%%%%%%%%%%%%%%%%%%%%%%%%%%%%%%%%%%

\section{General expressions}\label{appendixC}

In this Appendix, we establish the general expressions for the five contributions $\mathcal{P}_\ell(\omega)$ to the noise in terms of the self-energy and of the total double quantum dot Green function, respectively defined as
\begin{eqnarray}
 \mathbb{\Sigma}_{\alpha,ij}^{r,a,\lessgtr}(\varepsilon)=\sum_{k\in \alpha} V_{i\alpha}^*V_{j\alpha} g_{k\alpha}^{r,a,\lessgtr}(\varepsilon)~,
\end{eqnarray}
and
\begin{eqnarray}
 {\bf G}_{ij}^{r,a,\lessgtr}(\varepsilon)=\sum_{n\in i,m\in j }G_{in,jm}^{r,a,\lessgtr}(\varepsilon)~.
\end{eqnarray}

\subsection{Expression for $\mathcal{P}_1(\omega)$}

By starting from Eq.~(\ref{TFP1}), one gets
\begin{eqnarray}\label{resP1}
\mathcal{P}_1(\omega)&=& \frac{e^2}{h}\sum_{i,j=1,2} \int_{-\infty}^\infty d\varepsilon \Big[ \mathbb{\Sigma}_{\alpha,ji}^{<}(\varepsilon){\bf G}_{ij}^>(\varepsilon-\omega)\nonumber\\
&&+ \mathbb{\Sigma}_{\alpha,ji}^{>}(\varepsilon-\omega){\bf G}_{ij}^<(\varepsilon) \Big]\delta_{\alpha\beta}\nonumber\\
&=&  \frac{e^2}{h}\int_{-\infty}^\infty d\varepsilon \mathrm{Tr}\Big\{ \doubleunderline{\mathbb{\Sigma}_{\alpha}^{<}}(\varepsilon)\doubleunderline{{\bf G}^>}(\varepsilon-\omega)\nonumber\\
&&+\doubleunderline{ \mathbb{\Sigma}_{\alpha}^{>}}(\varepsilon-\omega)\doubleunderline{{\bf G}^<}(\varepsilon) \Big\}\delta_{\alpha\beta}~,
\end{eqnarray}
where $\mathrm{Tr}\{\}$ is the trace operator.

\subsection{Expression for $\mathcal{P}_2(\omega)$}

By starting from Eq.~(\ref{TFP2}), one gets
\begin{eqnarray}\label{resP2}
\mathcal{P}_2(\omega)&=&
-\frac{e^2}{\hbar} \sum_{i,j=1,2}\sum_{p,q=1,2} \int_{-\infty}^\infty d\varepsilon\nonumber\\
&&\times \Big[ {\bf G}_{jp}^{r}(\varepsilon)\mathbb{\Sigma}_{\alpha,pi}^{<}(\varepsilon)+{\bf G}_{jp}^{<}(\varepsilon)\mathbb{\Sigma}_{\alpha,pi}^{a}(\varepsilon) \Big] \nonumber\\
 &&\times
 \Big[ {\bf G}_{iq}^{>}(\varepsilon-\omega)\mathbb{\Sigma}_{\beta,qj}^{a}(\varepsilon-\omega)\nonumber\\
 &&+{\bf G}_{iq}^{r}(\varepsilon-\omega)\mathbb{\Sigma}_{\beta,qj}^{>}(\varepsilon-\omega) \Big]\nonumber\\
 &=&
-\frac{e^2}{\hbar}\int_{-\infty}^\infty d\varepsilon  \mathrm{Tr}\Big\{
 \Big[\doubleunderline{ {\bf G}^{r}}(\varepsilon)\doubleunderline{\mathbb{\Sigma}_{\alpha}^{<}}(\varepsilon)+\doubleunderline{{\bf G}^{<}}(\varepsilon)\doubleunderline{\mathbb{\Sigma}_{\alpha}^{a}}(\varepsilon) \Big] \nonumber\\
&&\times \Big[ \doubleunderline{{\bf G}^{>}}(\varepsilon-\omega)\doubleunderline{\mathbb{\Sigma}_{\beta}^{a}}(\varepsilon-\omega)\nonumber\\
&&+\doubleunderline{{\bf G}^{r}}(\varepsilon-\omega)\doubleunderline{\mathbb{\Sigma}_{\beta}^{>}}(\varepsilon-\omega) \Big]\Big\}~.
\end{eqnarray}

\subsection{Expression for $\mathcal{P}_3(\omega)$}

By starting from Eq.~(\ref{TFP3}), one gets
\begin{eqnarray}\label{resP3}
&&\mathcal{P}_3(\omega)=\frac{e^2}{h} \sum_{i,j=1,2}\sum_{p,q=1,2} \int_{-\infty}^\infty d\varepsilon {\bf G}_{ij}^>(\varepsilon-\omega)\nonumber \\
&&\times  \Big[ \mathbb{\Sigma}_{\beta ,jq}^{r}(\varepsilon){\bf G}_{pq}^{r}(\varepsilon)\mathbb{\Sigma}_{\alpha,pi}^{<}(\varepsilon)\nonumber \\
&&+\mathbb{\Sigma}_{\beta,jq}^{r}(\varepsilon){\bf G}_{pq}^{<}(\varepsilon)\mathbb{\Sigma}_{\alpha,pi}^{a}(\varepsilon)\nonumber\\
&&+\mathbb{\Sigma}_{\beta,jq}^{<}(\varepsilon){\bf G}_{pq}^{a}(\varepsilon)\mathbb{\Sigma}_{\alpha,pi}^{a}(\varepsilon) \Big]\nonumber \\
  &=&\frac{e^2}{h} \int_{-\infty}^\infty d\varepsilon \mathrm{Tr}\Big\{ \doubleunderline{{\bf G}^>}(\varepsilon-\omega)
  \Big[\doubleunderline{ \mathbb{\Sigma}_{\beta}^{r}}(\varepsilon)\doubleunderline{{\bf G}^{r}}(\varepsilon)\doubleunderline{\mathbb{\Sigma}_{\alpha}^{<}}(\varepsilon)\nonumber \\
  &&+\doubleunderline{\mathbb{\Sigma}_{\beta}^{r}}(\varepsilon)\doubleunderline{{\bf G}^{<}}(\varepsilon)\doubleunderline{\mathbb{\Sigma}_{\alpha}^{a}}(\varepsilon)
  +\doubleunderline{\mathbb{\Sigma}_{\beta}^{<}}(\varepsilon)\doubleunderline{{\bf G}^{a}}(\varepsilon)\doubleunderline{\mathbb{\Sigma}_{\alpha}^{a}}(\varepsilon) \Big]\Big\}~.
\end{eqnarray}

\subsection{Expression for $\mathcal{P}_4(\omega)$}

By starting from Eq.~(\ref{TFP4}), one gets
\begin{eqnarray}\label{resP4}
\mathcal{P}_4(\omega)&=&\frac{e^2}{h} \sum_{i,j=1,2}\sum_{p,q=1,2}
\int_{-\infty}^\infty d\varepsilon{\bf G}_{ji}^<(\varepsilon) \nonumber \\
&&\times\Big[ \mathbb{\Sigma}_{\alpha,ip}^{>}(\varepsilon-\omega){\bf G}_{pq}^{a}(\varepsilon-\omega)\mathbb{\Sigma}_{\beta,qj}^{a}(\varepsilon-\omega)\nonumber \\
 &&+\mathbb{\Sigma}_{\alpha,ip}^{r}(\varepsilon-\omega){\bf G}_{pn',qm'}^{>}(\varepsilon-\omega)\mathbb{\Sigma}_{\beta,qj}^{a}(\varepsilon-\omega)\nonumber \\
 &&+\mathbb{\Sigma}_{\alpha,ip}^{r}(\varepsilon-\omega){\bf G}_{pq}^{r}(\varepsilon-\omega)\mathbb{\Sigma}_{\beta,qj}^{>}(\varepsilon-\omega) \Big]\nonumber\\
 &=&\frac{e^2}{h}
\int_{-\infty}^\infty d\varepsilon \mathrm{Tr}\Big\{
\doubleunderline{{\bf G}^<}(\varepsilon) \nonumber\\
&&\times \Big[ \doubleunderline{\mathbb{\Sigma}_{\alpha}^{>}}(\varepsilon-\omega)\doubleunderline{{\bf G}^{a}}(\varepsilon-\omega)\doubleunderline{\mathbb{\Sigma}_{\beta}^{a}}(\varepsilon-\omega)\nonumber \\
 &&+\doubleunderline{\mathbb{\Sigma}_{\alpha}^{r}}(\varepsilon-\omega)\doubleunderline{{\bf G}^{>}}(\varepsilon-\omega)\doubleunderline{\mathbb{\Sigma}_{\beta}^{a}}(\varepsilon-\omega)\nonumber \\
 &&+\doubleunderline{\mathbb{\Sigma}_{\alpha}^{r}}(\varepsilon-\omega)\doubleunderline{{\bf G}^{r}}(\varepsilon-\omega)\doubleunderline{\mathbb{\Sigma}_{\beta}^{>}}(\varepsilon-\omega) \Big]\Big\}~.
\end{eqnarray}

\subsection{Expression for $\mathcal{P}_5(\omega)$}

By starting from Eq.~(\ref{TFP5}), one gets
\begin{eqnarray}\label{resP5}
\mathcal{P}_5(\omega)&=&-\frac{e^2}{h} \sum_{i,j=1,2}\sum_{p,q=1,2}
 \Big[ \mathbb{\Sigma}_{\alpha,ip}^{>}(\varepsilon-\omega){\bf G}_{pj}^{a}(\varepsilon-\omega)\nonumber \\
 &&+\mathbb{\Sigma}_{\alpha,ip}^{r}(\varepsilon-\omega){\bf G}_{pj}^{>}(\varepsilon-\omega) \Big]\nonumber \\
 &&\times\Big[ \mathbb{\Sigma}_{\beta,jq}^{r}(\varepsilon){\bf G}_{qi}^{<}(\varepsilon)+\mathbb{\Sigma}_{\beta,jq}^{<}(\varepsilon){\bf G}_{qi}^{a}(\varepsilon) \Big]\nonumber\\
 &=&-\frac{e^2}{h}
 \mathrm{Tr}\Big\{\Big[\doubleunderline{ \mathbb{\Sigma}_{\alpha}^{>}}(\varepsilon-\omega)\doubleunderline{{\bf G}^{a}}(\varepsilon-\omega)\nonumber \\
 &&+\doubleunderline{\mathbb{\Sigma}_{\alpha}^{r}}(\varepsilon-\omega)\doubleunderline{{\bf G}^{>}}(\varepsilon-\omega) \Big]\nonumber \\
 &&\times\Big[\doubleunderline{ \mathbb{\Sigma}_{\beta}^{r}}(\varepsilon)\doubleunderline{{\bf G}^{<}}(\varepsilon)+\doubleunderline{\mathbb{\Sigma}_{\beta}^{<}}(\varepsilon)\doubleunderline{{\bf G}^{a}}(\varepsilon) \Big]\Big\}~.
\end{eqnarray}

%%%%%%%%%%%%%%%%%%%%%%%%%%%%%%%%%%%%%%%%%%%%%%%%%%%%%%%
%                                                     %
%                                                     %
%          NOISE IN TERMS OF TRANSMISSION                                    %
%                                                     %
%                                                     %
%%%%%%%%%%%%%%%%%%%%%%%%%%%%%%%%%%%%%%%%%%%%%%%%%%%%%%%

\section{Finite-frequency noise in terms of transmission amplitude/coefficient}\label{appendixD}

The self-energy matrices associated with the reservoir~$\alpha$ are given by
\begin{eqnarray}
 &&\doubleunderline{\mathbb{\Sigma}_\alpha^{r,a}}(\varepsilon)=\mp\frac{i}{2} \doubleunderline{\Gamma_\alpha}~,\\
 &&\doubleunderline{\mathbb{\Sigma}_\alpha^<}(\varepsilon)=if_\alpha^e(\varepsilon) \doubleunderline{\Gamma_\alpha}~,\\
 &&\doubleunderline{\mathbb{\Sigma}_\alpha^>}(\varepsilon)=-if_\alpha^h(\varepsilon) \doubleunderline{\Gamma_\alpha}~,
\end{eqnarray}
where $f_\alpha^e(\varepsilon)=1/(1+\exp(\varepsilon-\mu_\alpha)/k_BT_\alpha)$ and $f_\alpha^h(\varepsilon) =1-f_\alpha^e(\varepsilon) $ are the Fermi-Dirac distribution functions for electrons and holes respectively, $T_\alpha$ is the temperature and $\mu_\alpha$ the chemical potential of the reservoir $\alpha$ with density of states $\rho_\alpha$, which is energy independent in the wide-band limit, and where $\Gamma_{\alpha,ij}=2\pi\rho_\alpha V^*_{i\alpha}V_{j\alpha}$ are the elements of the dot-reservoir coupling matrix $\doubleunderline{\Gamma_\alpha}$. Moreover,  one has
$\doubleunderline{{\bf G}^{\lessgtr}}(\varepsilon)= \sum_{\alpha=L,R}\doubleunderline{{\bf G}^r}(\varepsilon)\doubleunderline{\mathbb{\Sigma}^\lessgtr_\alpha}(\varepsilon)\doubleunderline{{\bf G}^{a}}(\varepsilon)$, and we define the transmission amplitude matrix, the transmission coefficient matrix and the effective transmission coefficient matrix as
\begin{eqnarray}
 &&\doubleunderline{t_{\alpha\alpha}}(\varepsilon)=i\doubleunderline{{\bf G}^r}(\varepsilon)\doubleunderline{\Gamma_\alpha}~,\nonumber\\
 &&\doubleunderline{\mathcal{T}_{\alpha\beta}}(\varepsilon)=\doubleunderline{{\bf G}^r}(\varepsilon)\,\doubleunderline{\Gamma_\alpha}\;\doubleunderline{{\bf G}^a}(\varepsilon)\,\doubleunderline{\Gamma_\beta}~,\nonumber\\
 &&\doubleunderline{\mathcal{T}_{\alpha\alpha}^\mathrm{eff}}(\varepsilon)=\doubleunderline{t_{\alpha\alpha}}(\varepsilon)+\doubleunderline{t_{\alpha\alpha}^+}(\varepsilon)-\doubleunderline{\mathcal{T}_{\alpha\alpha}}(\varepsilon)~,
\end{eqnarray}
where $\doubleunderline{t_{\alpha\alpha}^+}(\varepsilon)$ is the conjugate transpose of $\doubleunderline{t_{\alpha\alpha}}(\varepsilon)$.

\subsection{Expression for $\mathcal{P}_1(\omega)$}

By starting from Eq.~(\ref{resP1}), we have
\begin{eqnarray}
\mathcal{P}_1(\omega)
&=& \frac{ie^2}{h}\delta_{\alpha\beta}\int_{-\infty}^\infty d\varepsilon \mathrm{Tr}\Big\{ f_\alpha^e(\varepsilon) \doubleunderline{\Gamma_\alpha}\doubleunderline{{\bf G}^>}(\varepsilon-\omega)\nonumber \\
&&-f_\alpha^h(\varepsilon-\omega) \doubleunderline{\Gamma_\alpha}\doubleunderline{{\bf G}^<}(\varepsilon) \Big\}\nonumber\\
&&= \frac{ie^2}{h}\delta_{\alpha\beta}\sum_{\delta=L,R}\int_{-\infty}^\infty d\varepsilon \mathrm{Tr}\Big\{\nonumber\\
&&\times f_\alpha^e(\varepsilon) \doubleunderline{\Gamma_\alpha}\doubleunderline{{\bf G}^r}(\varepsilon-\omega)\doubleunderline{\mathbb{\Sigma}^>_\delta}(\varepsilon-\omega)\doubleunderline{{\bf G}^{a}}(\varepsilon-\omega)\nonumber \\
&&-f_\alpha^h(\varepsilon-\omega) \doubleunderline{\Gamma_\alpha}\doubleunderline{{\bf G}^r}(\varepsilon)\doubleunderline{\mathbb{\Sigma}^<_\delta}(\varepsilon)\doubleunderline{{\bf G}^{a}}(\varepsilon) \Big\}\nonumber\\
&&= \frac{e^2}{h}\delta_{\alpha\beta}\sum_{\delta=L,R}\int_{-\infty}^\infty d\varepsilon \mathrm{Tr}\Big\{\nonumber\\
&&\times f_\alpha^e(\varepsilon)f_\delta^h(\varepsilon-\omega) \doubleunderline{\Gamma_\alpha}\doubleunderline{{\bf G}^r}(\varepsilon-\omega) \doubleunderline{\Gamma_\delta}\doubleunderline{{\bf G}^{a}}(\varepsilon-\omega)\nonumber \\
&&+f_\alpha^h(\varepsilon-\omega)f_\delta^e(\varepsilon) \doubleunderline{\Gamma_\alpha}\doubleunderline{{\bf G}^r}(\varepsilon)\doubleunderline{\Gamma_\delta}\doubleunderline{{\bf G}^{a}}(\varepsilon) \Big\}~.
\end{eqnarray}

It leads to
\begin{eqnarray}\label{expP1}
\mathcal{P}_1(\omega)
&=& \frac{e^2}{h}\sum_{\delta=L,R}\int_{-\infty}^\infty d\varepsilon \mathrm{Tr}\Big\{ f_\alpha^e(\varepsilon)f_\delta^h(\varepsilon-\omega) \doubleunderline{\mathcal{T}_{\delta\alpha}}(\varepsilon-\omega)\nonumber \\
&&+f_\delta^e(\varepsilon) f_\alpha^h(\varepsilon-\omega)\doubleunderline{\mathcal{T}_{\delta\alpha}}(\varepsilon) \Big\}\delta_{\alpha\beta}~.
\end{eqnarray}

\subsection{Expression for $\mathcal{P}_2(\omega)$}

By starting from Eq.~(\ref{resP2}), we have
\begin{eqnarray}
&&\mathcal{P}_2(\omega)=
-\frac{e^2}{\hbar}\int_{-\infty}^\infty d\varepsilon  \mathrm{Tr}\Big\{
 \Big[\doubleunderline{ {\bf G}^{r}}(\varepsilon)\doubleunderline{\mathbb{\Sigma}_{\alpha}^{<}}(\varepsilon)\nonumber\\
 &&+\sum_{\gamma=L,R}\doubleunderline{{\bf G}^r}(\varepsilon)\doubleunderline{\mathbb{\Sigma}^<_\gamma}(\varepsilon)\doubleunderline{{\bf G}^{a}}(\varepsilon)\doubleunderline{\mathbb{\Sigma}_{\alpha}^{a}}(\varepsilon) \Big]\nonumber\\
 &&\times\Big[ \sum_{\delta=L,R}\doubleunderline{{\bf G}^r}(\varepsilon-\omega)\doubleunderline{\mathbb{\Sigma}^>_\delta}(\varepsilon-\omega)\doubleunderline{{\bf G}^{a}}(\varepsilon-\omega)\doubleunderline{\mathbb{\Sigma}_{\beta}^{a}}(\varepsilon-\omega)\nonumber\\
 &&+\doubleunderline{{\bf G}^{r}}(\varepsilon-\omega)\doubleunderline{\mathbb{\Sigma}_{\beta}^{>}}(\varepsilon-\omega) \Big]\Big\}\nonumber\\
 &&=
-\frac{e^2}{\hbar}\int_{-\infty}^\infty d\varepsilon  \mathrm{Tr}\Big\{
 \Big[f_\alpha^e(\varepsilon)\doubleunderline{ {\bf G}^{r}}(\varepsilon) \doubleunderline{\Gamma_\alpha}\nonumber\\
 &&+\frac{i}{2}\sum_{\gamma=L,R}f_\gamma^e(\varepsilon)\doubleunderline{{\bf G}^r}(\varepsilon) \doubleunderline{\Gamma_\gamma}\doubleunderline{{\bf G}^{a}}(\varepsilon)\doubleunderline{\Gamma_\alpha} \Big]\nonumber\\
 &&\times\Big[\frac{i}{2} \sum_{\delta=L,R}f_\delta^h(\varepsilon-\omega)\doubleunderline{{\bf G}^r}(\varepsilon-\omega) \doubleunderline{\Gamma_\delta}\doubleunderline{{\bf G}^{a}}(\varepsilon-\omega) \doubleunderline{\Gamma_\beta}\nonumber\\
&& +f_\beta^h(\varepsilon-\omega)\doubleunderline{{\bf G}^{r}}(\varepsilon-\omega) \doubleunderline{\Gamma_\beta}\Big]\Big\}~.
\end{eqnarray}

It leads to
\begin{eqnarray}\label{expP2}
\mathcal{P}_2(\omega)
&=&
\frac{e^2}{\hbar}\int_{-\infty}^\infty d\varepsilon  \mathrm{Tr}\Big\{
 \Big[\frac{1}{2} \sum_{\gamma=L,R}f_\gamma^e(\varepsilon)\doubleunderline{\mathcal{T}_{\gamma\alpha}}(\varepsilon)\nonumber\\
 &&-f_\alpha^e(\varepsilon)\doubleunderline{t_{\alpha\alpha}}(\varepsilon)  \Big]
\Big[\frac{1}{2} \sum_{\delta=L,R}f_\delta^h(\varepsilon-\omega)\doubleunderline{\mathcal{T}_{\delta\beta}}(\varepsilon-\omega)\nonumber\\
 &&-f_\beta^h(\varepsilon-\omega)\doubleunderline{t_{\beta\beta}}(\varepsilon-\omega) \Big]\Big\}~.
\end{eqnarray}

\subsection{Expression for $\mathcal{P}_3(\omega)$}

By starting from Eq.~(\ref{resP3}), we have
\begin{eqnarray}
\mathcal{P}_3(\omega)
  &=&\frac{e^2}{h} \int_{-\infty}^\infty d\varepsilon \mathrm{Tr}\Big\{ \nonumber\\
  &&\times\sum_{\gamma=L,R}\doubleunderline{{\bf G}^r}(\varepsilon-\omega)\doubleunderline{\mathbb{\Sigma}^>_\gamma}(\varepsilon-\omega)\doubleunderline{{\bf G}^{a}}(\varepsilon-\omega)\nonumber\\
  &&\times
  \Big[\doubleunderline{ \mathbb{\Sigma}_{\beta}^{r}}(\varepsilon)\doubleunderline{{\bf G}^{r}}(\varepsilon)\doubleunderline{\mathbb{\Sigma}_{\alpha}^{<}}(\varepsilon)\nonumber\\
&&  +\doubleunderline{\mathbb{\Sigma}_{\beta}^{r}}(\varepsilon)\sum_{\delta=L,R}\doubleunderline{{\bf G}^r}(\varepsilon)\doubleunderline{\mathbb{\Sigma}^<_\delta}(\varepsilon)\doubleunderline{{\bf G}^{a}}(\varepsilon)\doubleunderline{\mathbb{\Sigma}_{\alpha}^{a}}(\varepsilon)\nonumber\\
&&+\doubleunderline{\mathbb{\Sigma}_{\beta}^{<}}(\varepsilon)\doubleunderline{{\bf G}^{a}}(\varepsilon)\doubleunderline{\mathbb{\Sigma}_{\alpha}^{a}}(\varepsilon) \Big]\Big\}\nonumber\\
  &=&\frac{e^2}{h} \int_{-\infty}^\infty d\varepsilon \mathrm{Tr}\Big\{ \nonumber\\
  &&\times\sum_{\gamma=L,R}
  f_\gamma^h(\varepsilon-\omega)\doubleunderline{{\bf G}^r}(\varepsilon-\omega) \doubleunderline{\Gamma_\gamma}\doubleunderline{{\bf G}^{a}}(\varepsilon-\omega) \doubleunderline{\Gamma_\beta}\nonumber\\
  &&\times
  \Big[\frac{1}{4}\sum_{\delta=L,R} f_\delta^e(\varepsilon)\doubleunderline{{\bf G}^r}(\varepsilon) \doubleunderline{\Gamma_\delta}\doubleunderline{{\bf G}^{a}}(\varepsilon)\doubleunderline{\Gamma_\alpha}\nonumber\\
  &&-\frac{i}{2}  f_\alpha^e(\varepsilon)\doubleunderline{{\bf G}^{r}}(\varepsilon)\doubleunderline{\Gamma_\alpha}+\frac{i}{2} f_\beta^e(\varepsilon) \doubleunderline{{\bf G}^{a}}(\varepsilon) \doubleunderline{\Gamma_\alpha} \Big]\Big\}~.
\end{eqnarray}

It leads to
\begin{eqnarray}\label{expP3}
 \mathcal{P}_3(\omega)
  &=&\frac{e^2}{h} \int_{-\infty}^\infty d\varepsilon \mathrm{Tr}\Big\{  \sum_{\gamma=L,R}
  f_\gamma^h(\varepsilon-\omega)\doubleunderline{\mathcal{T}_{\gamma\beta}}(\varepsilon-\omega)\nonumber\\
&&\times  \Big[\frac{1}{4} \sum_{\delta=L,R} f_\delta^e(\varepsilon)\doubleunderline{\mathcal{T}_{\delta\alpha}}(\varepsilon)\nonumber\\
&&-\frac{1}{2}  f_\alpha^e(\varepsilon)\doubleunderline{t_{\alpha\alpha}}(\varepsilon)-\frac{1}{2} f_\beta^e(\varepsilon) \doubleunderline{t_{\alpha\alpha}^+}(\varepsilon)\Big]\Big\}~.
\end{eqnarray}

\subsection{Expression for $\mathcal{P}_4(\omega)$}

By starting from Eq.~(\ref{resP4}), we have
\begin{eqnarray}
\mathcal{P}_4(\omega)
  &=&\frac{e^2}{h}
\int_{-\infty}^\infty d\varepsilon \mathrm{Tr}\Big\{
\sum_{\gamma=L,R}\doubleunderline{{\bf G}^r}(\varepsilon)\doubleunderline{\mathbb{\Sigma}^<_\gamma}(\varepsilon)\doubleunderline{{\bf G}^{a}}(\varepsilon) \nonumber\\
&&\times\Big[ \doubleunderline{\mathbb{\Sigma}_{\alpha}^{>}}(\varepsilon-\omega)\doubleunderline{{\bf G}^{a}}(\varepsilon-\omega)\doubleunderline{\mathbb{\Sigma}_{\beta}^{a}}(\varepsilon-\omega)\nonumber \\
 &&+ \sum_{\delta=L,R}
 \doubleunderline{\mathbb{\Sigma}_{\alpha}^{r}}(\varepsilon-\omega)\doubleunderline{{\bf G}^r}(\varepsilon-\omega)\nonumber\\
 &&\times\doubleunderline{\mathbb{\Sigma}^>_\delta}(\varepsilon-\omega)\doubleunderline{{\bf G}^{a}}(\varepsilon-\omega)
 \doubleunderline{\mathbb{\Sigma}_{\beta}^{a}}(\varepsilon-\omega)\nonumber\\
&& +\doubleunderline{\mathbb{\Sigma}_{\alpha}^{r}}(\varepsilon-\omega)\doubleunderline{{\bf G}^{r}}(\varepsilon-\omega)\doubleunderline{\mathbb{\Sigma}_{\beta}^{>}}(\varepsilon-\omega) \Big]\Big\}\nonumber\\
  &=&\frac{e^2}{h}
\int_{-\infty}^\infty d\varepsilon \mathrm{Tr}\Big\{
\sum_{\gamma=L,R}f_\gamma^e(\varepsilon)\doubleunderline{{\bf G}^r}(\varepsilon) \doubleunderline{\Gamma_\gamma}\doubleunderline{{\bf G}^{a}}(\varepsilon)  \doubleunderline{\Gamma_\alpha}\nonumber \\
&&\times\Big[ \frac{i}{2}f_\alpha^h(\varepsilon-\omega)\doubleunderline{{\bf G}^{a}}(\varepsilon-\omega) \doubleunderline{\Gamma_\beta}\nonumber \\
 &&+\frac{1}{4} \sum_{\delta=L,R}f_\delta^h(\varepsilon-\omega)
\doubleunderline{{\bf G}^r}(\varepsilon-\omega) \doubleunderline{\Gamma_\delta}\doubleunderline{{\bf G}^{a}}(\varepsilon-\omega)
  \doubleunderline{\Gamma_\beta}\nonumber \\
  &&-\frac{i}{2}f_\beta^h(\varepsilon-\omega) \doubleunderline{{\bf G}^{r}}(\varepsilon-\omega) \doubleunderline{\Gamma_\beta} \Big]\Big\}~.
\end{eqnarray}

It leads to
\begin{eqnarray}\label{expP4}
\mathcal{P}_4(\omega)
  &=&\frac{e^2}{h}
\int_{-\infty}^\infty d\varepsilon \mathrm{Tr}\Big\{
\sum_{\gamma=L,R}f_\gamma^e(\varepsilon)\doubleunderline{\mathcal{T}_{\gamma\alpha}}(\varepsilon)\nonumber\\
&&\times\Big[ \frac{1}{4} \sum_{\delta=L,R}f_\delta^h(\varepsilon-\omega)
\doubleunderline{\mathcal{T}_{\delta\beta}}(\varepsilon-\omega) \nonumber \\
 &&-\frac{1}{2} f_\alpha^h(\varepsilon-\omega)\doubleunderline{t^+_{\beta\beta}}(\varepsilon-\omega)\nonumber\\
&& -\frac{1}{2} f_\beta^h(\varepsilon-\omega) \doubleunderline{t_{\beta\beta}}(\varepsilon-\omega)  \Big]\Big\}~.
\end{eqnarray}

\subsection{Expression for $\mathcal{P}_5(\omega)$}

By starting from Eq.~(\ref{resP5}), we have
\begin{eqnarray}
&&\mathcal{P}_5(\omega)
 =-\frac{e^2}{h}
 \mathrm{Tr}\Big\{\Big[\doubleunderline{ \mathbb{\Sigma}_{\alpha}^{>}}(\varepsilon-\omega)\doubleunderline{{\bf G}^{a}}(\varepsilon-\omega)\nonumber\\
 &&+\doubleunderline{\mathbb{\Sigma}_{\alpha}^{r}}(\varepsilon-\omega)\sum_{\gamma=L,R}\doubleunderline{{\bf G}^r}(\varepsilon-\omega)\doubleunderline{\mathbb{\Sigma}^>_\gamma}(\varepsilon-\omega)\doubleunderline{{\bf G}^{a}}(\varepsilon-\omega) \Big]\nonumber \\
 &&\times\Big[\doubleunderline{ \mathbb{\Sigma}_{\beta}^{r}}(\varepsilon) \sum_{\delta=L,R}\doubleunderline{{\bf G}^r}(\varepsilon)\doubleunderline{\mathbb{\Sigma}^<_\delta}(\varepsilon)\doubleunderline{{\bf G}^{a}}(\varepsilon)+\doubleunderline{\mathbb{\Sigma}_{\beta}^{<}}(\varepsilon)\doubleunderline{{\bf G}^{a}}(\varepsilon) \Big]\Big\}\nonumber \\
  &=&\frac{e^2}{h}
 \mathrm{Tr}\Big\{\Big[if_\alpha^h(\varepsilon-\omega) \doubleunderline{\Gamma_\alpha}\doubleunderline{{\bf G}^{a}}(\varepsilon-\omega)\nonumber\\
 &&+\frac{1}{2}  \sum_{\gamma=L,R}f_\gamma^h(\varepsilon-\omega)\doubleunderline{\Gamma_\alpha}\doubleunderline{{\bf G}^r}(\varepsilon-\omega) \doubleunderline{\Gamma_\gamma}\doubleunderline{{\bf G}^{a}}(\varepsilon-\omega) \Big]\nonumber \\
 &&\times\Big[\frac{1}{2}   \sum_{\delta=L,R}f_\delta^e(\varepsilon)\doubleunderline{\Gamma_\beta}\doubleunderline{{\bf G}^r}(\varepsilon) \doubleunderline{\Gamma_\delta}\doubleunderline{{\bf G}^{a}}(\varepsilon)+if_\beta^e(\varepsilon) \doubleunderline{\Gamma_\beta}\doubleunderline{{\bf G}^{a}}(\varepsilon) \Big]\Big\}\nonumber \\
  &=&\frac{e^2}{h}
 \mathrm{Tr}\Big\{\Big[if_\alpha^h(\varepsilon-\omega) \doubleunderline{{\bf G}^{a}}(\varepsilon-\omega)\doubleunderline{\Gamma_\beta}\nonumber \\
&& +\frac{1}{2}  \sum_{\gamma=L,R}f_\gamma^h(\varepsilon-\omega)\doubleunderline{{\bf G}^r}(\varepsilon-\omega) \doubleunderline{\Gamma_\gamma}\doubleunderline{{\bf G}^{a}}(\varepsilon-\omega) \doubleunderline{\Gamma_\beta}\Big]\nonumber \\
 &&\times\Big[\frac{1}{2}   \sum_{\delta=L,R}f_\delta^e(\varepsilon)\doubleunderline{{\bf G}^r}(\varepsilon) \doubleunderline{\Gamma_\delta}\doubleunderline{{\bf G}^{a}}(\varepsilon)\doubleunderline{\Gamma_\alpha}
 +if_\beta^e(\varepsilon) \doubleunderline{{\bf G}^{a}}(\varepsilon)\doubleunderline{\Gamma_\alpha} \Big]\Big\}~.\nonumber \\
\end{eqnarray}

It leads to
\begin{eqnarray}\label{expP5}
\mathcal{P}_5(\omega)
&=&\frac{e^2}{h}
 \mathrm{Tr}\Big\{\Big[ \frac{1}{2}  \sum_{\gamma=L,R}f_\gamma^h(\varepsilon-\omega)\doubleunderline{\mathcal{T}_{\gamma\beta}}(\varepsilon-\omega)\nonumber \\
 &&-f_\alpha^h(\varepsilon-\omega) \doubleunderline{t_{\beta\beta}^+}(\varepsilon-\omega)\Big]\nonumber\\
&&\times\Big[\frac{1}{2}  \sum_{\delta=L,R}f_\delta^e(\varepsilon)\doubleunderline{\mathcal{T}_{\delta\alpha}}(\varepsilon)
 -f_\beta^e(\varepsilon) \doubleunderline{t_{\alpha\alpha}^+}(\varepsilon) \Big]\Big\}~.\nonumber \\
\end{eqnarray}

%%%%%%%%%%%%%%%%%%%%%%%%%%%%%%%%%%%%%%%%%%%%%%%%%%%%%%%
%                                                     %
%                                                     %
%          FINAL RESULT                               %
%                                                     %
%                                                     %
%%%%%%%%%%%%%%%%%%%%%%%%%%%%%%%%%%%%%%%%%%%%%%%%%%%%%%%

\section{Final result}\label{appendixE}

By collecting Eqs.~(\ref{expP1}), (\ref{expP2}), (\ref{expP3}), (\ref{expP4}) and (\ref{expP5}), one finally obtains for the finite-frequency noise
\begin{eqnarray}
\mathcal{S}_{\alpha\beta}(\omega)&=&
\frac{e^2}{h}\int_{-\infty}^\infty d\varepsilon \mathrm{Tr}\Big\{\nonumber\\
&&\times\sum_{\delta=L,R}f_\alpha^e(\varepsilon)f_\delta^h(\varepsilon-\omega) \doubleunderline{\mathcal{T}_{\delta\alpha}}(\varepsilon-\omega)\nonumber\\
&&+\sum_{\delta=L,R}f_\delta^e(\varepsilon) f_\alpha^h(\varepsilon-\omega)\doubleunderline{\mathcal{T}_{\delta\alpha}}(\varepsilon) \Big\}\delta_{\alpha\beta}
\nonumber \\
&&+\frac{e^2}{\hbar}\int_{-\infty}^\infty d\varepsilon  \mathrm{Tr}\Big\{\nonumber\\
&&\times \Big[\frac{1}{2}\sum_{\gamma=L,R}f_\gamma^e(\varepsilon)\doubleunderline{\mathcal{T}_{\gamma\alpha}}(\varepsilon)-f_\alpha^e(\varepsilon)\doubleunderline{t_{\alpha\alpha}}(\varepsilon)  \Big]\nonumber\\
&&\times\Big[\frac{1}{2}\sum_{\delta=L,R}f_\delta^h(\varepsilon-\omega)\doubleunderline{\mathcal{T}_{\delta\beta}}(\varepsilon-\omega)\nonumber\\
&& -f_\beta^h(\varepsilon-\omega)\doubleunderline{t_{\beta\beta}}(\varepsilon-\omega) \Big]\Big\}
\nonumber \\
&&+\frac{e^2}{h} \int_{-\infty}^\infty d\varepsilon \mathrm{Tr}\Big\{  \sum_{\gamma=L,R}
  f_\gamma^h(\varepsilon-\omega)\doubleunderline{\mathcal{T}_{\gamma\beta}}(\varepsilon-\omega)\nonumber\\
&&\times  \Big[\frac{1}{4}\sum_{\delta=L,R} f_\delta^e(\varepsilon)\doubleunderline{\mathcal{T}_{\delta\alpha}}(\varepsilon)\nonumber\\
 && -\frac{1}{2} f_\alpha^e(\varepsilon)\doubleunderline{t_{\alpha\alpha}}(\varepsilon)
  -\frac{1}{2}f_\beta^e(\varepsilon) \doubleunderline{t_{\alpha\alpha}^+}(\varepsilon)\Big]\Big\}
\nonumber \\
&&+\frac{e^2}{h}
\int_{-\infty}^\infty d\varepsilon \mathrm{Tr}\Big\{
\sum_{\gamma=L,R}f_\gamma^e(\varepsilon)\doubleunderline{\mathcal{T}_{\gamma\alpha}}(\varepsilon)\nonumber\\
&&\times\Big[\frac{1}{4}\sum_{\delta=L,R}f_\delta^h(\varepsilon-\omega)
\doubleunderline{\mathcal{T}_{\delta\beta}}(\varepsilon-\omega)\nonumber \\
&&-\frac{1}{2}f_\alpha^h(\varepsilon-\omega)\doubleunderline{t^+_{\beta\beta}}(\varepsilon-\omega)\nonumber\\
&& -\frac{1}{2}f_\beta^h(\varepsilon-\omega) \doubleunderline{t_{\beta\beta}}(\varepsilon-\omega)  \Big]\Big\}\nonumber \\
&&+\frac{e^2}{h}
 \mathrm{Tr}\Big\{\Big[ \frac{1}{2} \sum_{\gamma=L,R}f_\gamma^h(\varepsilon-\omega)\doubleunderline{\mathcal{T}_{\gamma\beta}}(\varepsilon-\omega)\nonumber \\
 &&-f_\alpha^h(\varepsilon-\omega) \doubleunderline{t_{\beta\beta}^+}(\varepsilon-\omega)
\Big]\nonumber\\
&&\times\Big[\frac{1}{2}  \sum_{\delta=L,R}f_\delta^e(\varepsilon)\doubleunderline{\mathcal{T}_{\delta\alpha}}(\varepsilon)
 -f_\beta^e(\varepsilon) \doubleunderline{t_{\alpha\alpha}^+}(\varepsilon) \Big]\Big\}~.\nonumber\\
\end{eqnarray}
From this general expression, we can explicitly express the auto-correlators and the cross-correlators by replacing the indices $\alpha$ and $\beta$ by $L$ or $R$, as done below.

\subsection{Auto-correlator $\mathcal{S}_{LL}(\omega)$}

\begin{eqnarray}\label{appendixSLL}
 &&\mathcal{S}_{LL}(\omega)=\frac{e^2}{h}\int_{-\infty}^{\infty}d\varepsilon\, \mathrm{Tr} \bigg\{f^e_L(\varepsilon)f^h_L(\varepsilon-\omega)\nonumber\\
&&\times\left[\doubleunderline{\mathcal{T}_{LL}^\mathrm{eff}}(\varepsilon)\doubleunderline{\mathcal{T}_{LL}^\mathrm{eff}}(\varepsilon-\omega)\right.\nonumber\\
&& +\left.\left(\doubleunderline{t_{LL}}(\varepsilon)-\doubleunderline{t_{LL}}(\varepsilon-\omega)\right)\left(\doubleunderline{t^+_{LL}}(\varepsilon)-\doubleunderline{t^+_{LL}}(\varepsilon-\omega)\right)  \right]\nonumber\\
 && +f^e_R(\varepsilon)f^h_R(\varepsilon-\omega)\doubleunderline{\mathcal{T}_{RL}}(\varepsilon)\doubleunderline{\mathcal{T}_{RL}}(\varepsilon-\omega)\nonumber\\
&&+f^e_L(\varepsilon)f^h_R(\varepsilon-\omega)\left[1-\doubleunderline{\mathcal{T}_{LL}^\mathrm{eff}}(\varepsilon)\right]\doubleunderline{\mathcal{T}_{RL}}(\varepsilon-\omega)
\nonumber\\
 && +f^e_R(\varepsilon)f^h_L(\varepsilon-\omega)\doubleunderline{\mathcal{T}_{RL}}(\varepsilon)\left[1-\doubleunderline{\mathcal{T}_{LL}^\mathrm{eff}}(\varepsilon-\omega)\right] \bigg\}~.
\end{eqnarray}

\subsection{Auto-correlator $\mathcal{S}_{RR}(\omega)$}

\begin{eqnarray}\label{appendixSRR}
 &&\mathcal{S}_{RR}(\omega)=\frac{e^2}{h}\int_{-\infty}^{\infty}d\varepsilon\, \mathrm{Tr} \bigg\{f^e_R(\varepsilon)f^h_R(\varepsilon-\omega)\nonumber\\
&&\times\left[\doubleunderline{\mathcal{T}_{RR}^\mathrm{eff}}(\varepsilon)\doubleunderline{\mathcal{T}_{RR}^\mathrm{eff}}(\varepsilon-\omega)\right.\nonumber\\
&& +\left.\left(\doubleunderline{t_{RR}}(\varepsilon)-\doubleunderline{t_{RR}}(\varepsilon-\omega)\right)\left(\doubleunderline{t^+_{RR}}(\varepsilon)-\doubleunderline{t^+_{RR}}(\varepsilon-\omega)\right)  \right]\nonumber\\
 && +f^e_L(\varepsilon)f^h_L(\varepsilon-\omega)\doubleunderline{\mathcal{T}_{LR}}(\varepsilon)\doubleunderline{\mathcal{T}_{LR}}(\varepsilon-\omega)\nonumber\\
&&+f^e_R(\varepsilon)f^h_L(\varepsilon-\omega)\left[1-\doubleunderline{\mathcal{T}_{RR}^\mathrm{eff}}(\varepsilon)\right]\doubleunderline{\mathcal{T}_{LR}}(\varepsilon-\omega)
\nonumber\\
 && +f^e_L(\varepsilon)f^h_R(\varepsilon-\omega)\doubleunderline{\mathcal{T}_{LR}}(\varepsilon)\left[1-\doubleunderline{\mathcal{T}_{RR}^\mathrm{eff}}(\varepsilon-\omega)\right] \bigg\}~.
\end{eqnarray}

\subsection{Cross-correlator $\mathcal{S}_{LR}(\omega)$}

 \begin{eqnarray}\label{appendixSLR}
&& \mathcal{S}_{LR}(\omega)=\frac{e^2}{h}\int_{-\infty}^{\infty}d\varepsilon\, \mathrm{Tr} \bigg\{f^e_L(\varepsilon)f^h_L(\varepsilon-\omega)\nonumber\\
&&\times\left[\left[\doubleunderline{\mathcal{T}_{LL}}(\varepsilon)-\doubleunderline{t_{LL}}(\varepsilon)\right]\doubleunderline{\mathcal{T}_{LR}}(\varepsilon-\omega)
  -\doubleunderline{\mathcal{T}_{LL}}(\varepsilon)\doubleunderline{t_{RR}^+}(\varepsilon-\omega)\right]\nonumber\\
 && +f^e_R(\varepsilon)f^h_R(\varepsilon-\omega)\nonumber\\
 &&\times
 \left[\left[\doubleunderline{\mathcal{T}_{RL}}(\varepsilon)-\doubleunderline{t_{LL}^+}(\varepsilon)\right]\doubleunderline{\mathcal{T}_{RR}}(\varepsilon-\omega)
  -\doubleunderline{\mathcal{T}_{RL}}(\varepsilon)\doubleunderline{t_{RR}}(\varepsilon-\omega)\right]
\nonumber\\
 && +f^e_L(\varepsilon)f^h_R(\varepsilon-\omega)  \left[\doubleunderline{\mathcal{T}_{LL}}(\varepsilon)-\doubleunderline{t_{LL}}(\varepsilon)\right]\nonumber\\
 &&\times
  \left[\doubleunderline{\mathcal{T}_{RR}}(\varepsilon-\omega)-\doubleunderline{t_{RR}}(\varepsilon-\omega)\right]
\nonumber\\
 && +f^e_R(\varepsilon)f^h_L(\varepsilon-\omega)\left[\doubleunderline{\mathcal{T}_{RL}}(\varepsilon)-\doubleunderline{t_{LL}^+}(\varepsilon)\right]\nonumber\\
 &&\times
     \left[\doubleunderline{\mathcal{T}_{LR}}(\varepsilon-\omega)-\doubleunderline{t_{RR}^+}(\varepsilon-\omega)\right]
 \bigg\}~.
\end{eqnarray}

\subsection{Cross-correlator $\mathcal{S}_{RL}(\omega)$}

 \begin{eqnarray}\label{appendixSRL}
&& \mathcal{S}_{RL}(\omega)=\frac{e^2}{h}\int_{-\infty}^{\infty}d\varepsilon\, \mathrm{Tr} \bigg\{f^e_R(\varepsilon)f^h_R(\varepsilon-\omega)\nonumber\\
&&\times\left[\left[\doubleunderline{\mathcal{T}_{RR}}(\varepsilon)-\doubleunderline{t_{RR}}(\varepsilon)\right]\doubleunderline{\mathcal{T}_{RL}}(\varepsilon-\omega)
  -\doubleunderline{\mathcal{T}_{RR}}(\varepsilon)\doubleunderline{t_{LL}^+}(\varepsilon-\omega)\right]\nonumber\\
 && +f^e_L(\varepsilon)f^h_L(\varepsilon-\omega)\nonumber\\
 &&\times
 \left[\left[\doubleunderline{\mathcal{T}_{LR}}(\varepsilon)-\doubleunderline{t_{RR}^+}(\varepsilon)\right]\doubleunderline{\mathcal{T}_{LL}}(\varepsilon-\omega)
  -\doubleunderline{\mathcal{T}_{LR}}(\varepsilon)\doubleunderline{t_{LL}}(\varepsilon-\omega)\right]
\nonumber\\
 && +f^e_R(\varepsilon)f^h_L(\varepsilon-\omega)  \left[\doubleunderline{\mathcal{T}_{RR}}(\varepsilon)-\doubleunderline{t_{RR}}(\varepsilon)\right]\nonumber\\
 &&\times
  \left[\doubleunderline{\mathcal{T}_{LL}}(\varepsilon-\omega)-\doubleunderline{t_{LL}}(\varepsilon-\omega)\right]
\nonumber\\
 && +f^e_L(\varepsilon)f^h_R(\varepsilon-\omega)\left[\doubleunderline{\mathcal{T}_{LR}}(\varepsilon)-\doubleunderline{t_{RR}^+}(\varepsilon)\right]\nonumber\\
 &&\times
     \left[\doubleunderline{\mathcal{T}_{RL}}(\varepsilon-\omega)-\doubleunderline{t_{LL}^+}(\varepsilon-\omega)\right]
 \bigg\}~.
\end{eqnarray}

%%%%%%%%%%%%%%%%%%%%%%%%%%%%%%%%%%%%%%%%%%%%%%%%%%%%%%%%%%%%%%%%%%
%																 %
%																 %
%		REFERENCES											     %
%																 %
%																 %
%%%%%%%%%%%%%%%%%%%%%%%%%%%%%%%%%%%%%%%%%%%%%%%%%%%%%%%%%%%%%%%%%%

\bibliographystyle{iopart-num}
\bibliography{DQD_Noise_bibliography}

\end{document}